\documentclass[apj]{emulateapj}
\journalinfo{Accepted by ApJ on July 8, 2007}
\submitted{Submitted April 26, 2007 \hspace{1ex} Accepted July 8, 2007}

\newcommand{\lya}{Ly$\alpha$}
\newcommand{\fet}{Fe\,{\scriptsize II}}
\newcommand{\mgt}{Mg\,{\scriptsize II}}
\newcommand{\ct}{C\,{\scriptsize IV}}

\shorttitle{Black hole masses and enrichment of $z \sim 6$ SDSS quasars}
\shortauthors{Kurk et al.}

\begin{document}


\title{Black hole masses and enrichment of $z \sim 6$ SDSS quasars\altaffilmark{1}}


\author{Jaron D. Kurk\altaffilmark{2}, Fabian Walter}
\affil{Max--Planck--Institut f\"ur Astronomie, K\"onigstuhl 17, 
             D--69117, Heidelberg, Germany}

\author{Xiaohui Fan, Linhua Jiang}
\affil{Steward Observatory, The University of Arizona,
             933 N. Cherry Avenue, Tucson, AZ 85721}

\author{Dominik A. Riechers, Hans-Walter Rix}
\affil{Max--Planck--Institut f\"ur Astronomie, K\"onigstuhl 17, 
             D--69117, Heidelberg, Germany}

\author{Laura Pentericci}
\affil{Osservatorio Astronomico di Roma, Via di Frascati 33,
              I--00040, Monte Porzio Catone, Italy}

\author{Michael A. Strauss}
\affil{Department of Astrophysical Sciences, Princeton University, 
             Peyton Hall - Ivy Lane, Princeton, NJ 08544}

\author{Chris Carilli}
\affil{National Radio Astronomy Observatory, PO Box O, Socorro, NM 87801}

\and

\author{Stefan Wagner}
\affil{Landessternwarte Heidelberg, K\"onigstuhl 12,
             D--69117, Heidelberg, Germany}

\altaffiltext{1}{Based on observations carried out at the European Southern
  Observatory, Paranal, Chile under program Nos.\ 267.A--5689, 069.B--0289,
  071.B--0525, 074.A--0447 and 076.A--0304.}
\altaffiltext{2}{German Science Foundation's Collaborative Research Center 439 fellow.}

\begin{abstract}
  
  We present sensitive near--infrared spectroscopic observations for a sample
  of five $z \sim 6$ quasars. These quasars are amongst the most distant,
  currently known quasars in the universe.  The spectra have been obtained
  using ISAAC at the VLT and include the \ion{C}{4}, \ion{Mg}{2} and
  \ion{Fe}{2} lines, which are powerful probes of the chemical enrichment and
  the black hole masses in these objects.  We measure the
  \ion{Fe}{2}/\ion{Mg}{2} line ratio, as an observational proxy for the
  Fe/$\alpha$ element ratio.  These elements are produced by different types
  of supernovae and their abundance ratio can therefore serve as a
  cosmological clock.  We derive a ratio of 2.7$\pm$0.8 for our sample, which
  is similar to that found for lower redshift quasars, i.e., we provide
  additional evidence for the lack of evolution in the \ion{Fe}{2}/\ion{Mg}{2}
  line ratio of quasars up to the highest redshifts.  This result demonstrates
  that the sample quasars must have undergone a major episode of iron
  enrichment in less than one Gyr and star formation must have commenced at $z
  \ge 8$.  The linewidths of the \ion{Mg}{2} and \ion{C}{4} lines give two
  estimates for the black hole masses.  A third estimate is given by assuming
  that the quasars emit at their Eddington luminosity.  The derived masses
  using these three methods agree well, implying that the quasars are not
  likely to be strongly lensed.  We derive central black hole masses of $0.3 -
  5.2 \times 10^9$\,M$_\odot$.  These include the lowest black hole mass ever
  measured at $z \sim 6$, suggesting that we probe a more typical quasar
  population (with lower masses/luminosities) than examined before.  We use
  the difference between the redshift of \ion{Mg}{2} (a proxy for the systemic
  redshift of the quasar) and the onset of the Gunn Peterson trough to derive
  the extent of the ionized Str\"omgren spheres around our target quasars.
  The derived physical radii are about five Mpc.  Using a simple ionization
  model, the emission of the central quasars would need of order
  $10^6-10^8$\,yr to create these cavities in a surrounding intergalactic
  medium (IGM) with a neutral fraction between 0.1 and 1.0.  As the e--folding
  time scale for the central accreting black hole is on the order of a few
  times 10$^7$ years, it can grow by one e--folding or less within this time
  span.

\end{abstract}



\section{Introduction}\label{sec:introduction}

Since the advent of the Sloan Digital Sky Survey \citep[SDSS,][]{yor00} and
other large multi--band surveys, more than twenty quasars at $z \geq 5.7$ have
been discovered (mostly by the SDSS: \citealt{fan00,fan01,fan03,fan04,fan06a},
and three in other surveys: \citealt{got06,mcg06,ven07}).  These quasars
provide the best probes of the early growth of supermassive black holes and
serve as signposts for overdensities in the universe only one Gyr after the
Big Bang.  High redshift QSO studies also provide important insight into the
reionization \citep{fan06d} of the intergalactic medium (IGM) that took place
in the early Universe and into the relation between the formation of early
galaxies and black holes \citep{fan06b}.

Studies of the full SDSS quasar sample from $z = 0$ to $z = 6$
\citep{ric02a,sch05} show a strong evolution in number density of the quasar
phenomenon with redshift \citep[the comoving density of luminous quasars at $z
\sim 6$ is about forty times smaller than at $z \sim 3$,][]{ric06b}, but there
is no evolution detected in the intrinsic spectral properties of these quasar
phases \citep{fan06b}.  The high density peaks in the dark matter
distribution, which host QSOs, are expected to be much rarer in the early
universe \citep{she07}, while the immediate environment of quasars matures
apparently very early on.  Blueward of the Ly$\alpha$ line, the observed
spectra of SDSS quasars do change strongly as a function of redshift, as the
absorption by intervening neutral hydrogen increases with redshift up to the
point where the continuum emission is completely absorbed \citep[complete
Gunn--Peterson (GP) absorption,][]{gun65}.  The detection of deep GP troughs in
23 $z \sim 6$ QSOs \citep{fan06a} indicates that the IGM is significantly
neutral ($n_{\rm HI}/n_{\rm H} \sim 10^{-3} - 10^{-2}$) at these redshifts.
However, the large dispersion of IGM properties measured along different lines
of sight strongly suggests that the reionization process is more complex than
a quick phase transition over a narrow redshift range \citep{fan06c}. This
finding is supported by the results of the Wilkinson Microwave Anisotropy
Probe \citep[WMAP,][]{ben03} mission, which suggest that the universe may have
been $\sim$ 50\% neutral at redshift $z\gtrsim10$, but reionization did not
start before $z \sim 14$ \citep{pag07}.  QSOs acting alone cannot have
reionized the hydrogen in the IGM prior to $z \sim 4$ \citep[e.g.,][]{mei05},
leaving star--forming galaxies as the most likely candidates to ionize the IGM
at $z > 6$.

The strong emission lines excited by the nuclear source of a galaxy can be
used to infer various physical and dynamical properties of the circumnuclear
gas and therefore its power source, the nuclear black hole.  For QSOs at $z >
5.7$, the strong emission lines redward of Ly$\alpha$ are shifted into the
near infrared.  Several near--infrared (NIR) spectroscopy studies of QSO
samples including objects at $z \sim 6$ have been carried out
\citep[e.g.,][]{mai01, mai03, aok02, pen02, bar03, die03b, fre03, wil03,
  iwa04}.  The results of these early studies show that the
\ion{Fe}{2}/\ion{Mg}{2} line ratio appears to be similar at all redshifts, up
to and including $z \sim 6$ (some studies even report higher values than found
in the local universe).  Assuming that this ratio is directly related to the
Fe/$\alpha$ abundance, this suggests that all observed quasars have undergone
a major episode of iron enrichment early on, at least in their nuclear region.
We do note one caveat here: at high redshifts, we do not know about the
metallicity of less luminous QSOs as, so far, only the brightest objects have
been studied.

The Fe to Mg abundance ratio, and its observational proxy, the
\ion{Fe}{2}/\ion{Mg}{2} line ratio, can be considered a cosmological clock.
Both elements are produced in supernova explosions, but while Fe is produced
by Type Ia supernovae (SNe), which have relatively low mass progenitors (white
dwarfs in binary systems), Mg is produced by Type II SNe, which have high mass
progenitors.  Mg therefore appears almost instanteneously after initial star
formation while the Fe production starts only later.  The ratio of Fe to Mg is
predicted to build up quickly in the first 1 to 3 Gyr and then level off to
the value presently observed in the solar neighbourhood
\citep[e.g.,][]{yos98}.  More recent studies, however, show that this may be
true for our Milky Way and similar galaxies, but can be as short as 0.3\,Gyr
for elliptical galaxies \citep{fri98,mat01}, as the time of maximum enrichment
is a strong function of the adopted stellar life times, initial mass function,
and star formation rate.

Evidence for enrichment on kiloparsec scales is also available from the
detection of spatially resolved carbon, oxygen and dust in the interstellar
medium around a QSO at $z = 6.4$ \citep{wal03,ber03b,wal04,mai05}.
\citet{sim06} and \citet{web06} have observed two $z \sim 6$ QSOs in the NIR
at moderately high spectral resolution and have found several intervening
\ion{C}{4} absorption systems, suggesting that $\Omega_{\rm CIV}$ does not
evolve with redshift.  This therefore indicates that even at Mpc scales a
large fraction of intergalactic metals may already have been in place at $z >
6$.

As the width of the observed emission lines is believed to represent (at least
in part) ordered motion of gas in the very center of the quasars, these can be
employed to estimate the mass which drives this motion \citep[see, for
example,][]{mcl04,ves06}.  The width of the \ion{Mg}{2} line is used by
several authors to derive masses in the range $1-6 \times 10^9$ M$_\odot$ for
the central black hole in distant QSOs.  For the redshift record holder,
quasar SDSS J1148+5251 at $z = 6.42$, this indeed indicates that the quasar is
accreting at the maximal allowable rate for a black hole, adopting the
Eddington limit criterion \citep{wil03}.  This quasar has also been detected
at mm wavelengths, providing important information about its dust content
\citep{ber03b}, the mass of the H$_2$ gas reservoir in the host galaxy
\citep{wal03}, the density and temperature of CO gas \citep{ber03a} and its
dynamical mass \citep{wal04}.  The estimated dynamical mass of $\sim 5 \times
10^{10}$ M$_\odot$ is inconsistent with a $\sim 10^{12}$ M$_\odot$ stellar
bulge predicted if the M$_{\rm BH} - \sigma_{\rm bulge}$ relation \citep[which
seems to be valid at lower redshifts, e.g.,][]{geb00,fer00} were to hold at
$z \sim 6$.  This indicates that black holes may form prior to the assembly of
stellar bulges \citep{wal04}.  The detection of [\ion{C}{2}\,158$\mu$m]
emission in this QSO by \citet{mai05} shows that its host is undergoing an
intense burst of star formation.  \citet{wan07} present a sample of eighteen
$z \sim 6$ QSOs observed at submm wavelengths.  Eight of these are detected at
mJy sensitivities, strengthening the idea that massive starbursts exist in
quasar host galaxies.

Some of the above mentioned NIR studies have been hampered by low S/N and/or
limited wavelength coverage of the obtained spectra, or included only very few
$z \sim 6$ QSOs.  With the goal of measuring the physical properties from
high--quality observations of a larger number of $z \sim 6$ quasars in a
consistent way, we have observed a sample including all published QSOs at
redshift $z > 5.8$ and below declination $\delta = +15^{\rm \circ}$, i.e.,
those accessible from the four 8m apertures that constitute the Very Large
Telescope (VLT).  Our NIR ISAAC spectroscopy in several bands from 1.0 to
2.5\,$\mu$m covers the \ion{C}{4}, \ion{Mg}{2} and \ion{Fe}{2} lines.
The sample includes objects with lower luminosities than observed in earlier
studies and thereby helps to constrain the properties of presumably more
\emph{typical} QSOs at these redshifts.

This paper is divided in the following parts: in Sec.\ \ref{sec:obs} we
describe the sample, the observations and the data reduction we have carried
out. In Sec.\ \ref{sec:results}, the results are shown, including an
explanation of the fitting procedure, followed by the measurement of the
properties of the spectral features. In Sec.\ \ref{sec:discussion}, the
interpretation of these results is presented, amongst others, in terms of
metal enrichment and black hole mass.  Finally, we briefly summarize our
results in Sec.\ \ref{sec:summary}.  We assume the following $\Lambda$CDM
cosmology throughout the paper: $H_0 = 70$ km s$^{-1}$ Mpc$^{-1}$,
$\Omega_{\rm m} = 0.3$, $\Omega_{\rm b} = 0.04$ and $\Omega_\Lambda = 0.7$.
These parameters are also used in papers written by \citet{mcl04} and
\citet{ves06}, from which we use several relations.

\section{Sample definition, observations and data reduction}
\label{sec:obs}

Of the eleven quasars at $z > 5.8$ known in 2003, we have selected those that
are easily observable from the Southern Hemisphere, i.e., at declinations
lower than 15$^{\rm \circ}$.  This resulted in a sample of five targets with
magnitudes $18.7 < {\rm z^*_{AB}} < 20.5$ and redshifts from discovery papers
in the range $5.8 < z < 6.3$ (see Table \ref{tbl:objects} for names,
redshifts, magnitudes and discovery papers).

\begin{deluxetable*}{llrrrrrr}
\tabletypesize{\scriptsize}
\tablecaption{VLT ISAAC Observations\label{tbl:objects}}
\tablewidth{0pt}
\tablehead{
\colhead{Full name\tablenotemark{a}} &
\colhead{Ref\tablenotemark{b}} & 
\colhead{$z$} & 
\colhead{z$^{\rm*,c}_{\rm AB}$} &
\colhead{t$_{Z}^{\rm exp,d}$} & \colhead{t$_{J}^{\rm exp,d}$} & 
\colhead{t$_{H}^{\rm exp,d}$} & \colhead{t$_{K}^{\rm exp,d}$} \\
& & & & \multicolumn{1}{c}{[h]} & \multicolumn{1}{c}{[h]} & 
\multicolumn{1}{c}{[h]} & \multicolumn{1}{c}{[h]}
}
\startdata
SDSS J083643.85 +   005453.3 &(2)&5.82&18.74& 1.7 &--& 3.2 & 4.0 \\
SDSS J000552.34 $-$ 000655.8 &(3)&5.85&20.54& 3.3 &--&  -- & 5.3 \\
SDSS J141111.29 +   121737.4 &(3)&5.95&19.64& 2.0 &--&  -- & 6.0 \\
SDSS J130608.26 +   035626.3 &(2)&5.99&19.47& 1.6 &--&  -- & 4.6 \\
SDSS J103027.10 +   052455.0 &(2)&6.28&20.05&--& 1.6 &  -- &12.0
\enddata
\tablenotetext{a}{Full names are listed here. In the main text and following
  tables, only appropriate abbreviations are used, e.g., J0836+0054 for
  SDSS J083643.85 + 005453.3.}  
\tablenotetext{b}{Discovery papers, from which redshifts and
  magnitudes presented in the table have been taken: (1)
  \citealt{fan00}, (2) \citealt{fan01}, (3) \citealt{fan04}.}
\tablenotetext{c}{Magnitudes in $z'$--band from discovery papers in
  asinh AB system. Magnitude errors are between 0.05 and 0.10.}
\tablenotetext{d}{Exposures time in hours for the bands indicated.}
\end{deluxetable*}

The observations were carried out with the ISAAC instrument \citep{moo98} on
Antu (VLT--UT1) in low resolution (LR) mode, using the 1024$\times$1024 Hawaii
Rockwell array of the \emph{Short Wavelength} arm.  ISAAC's LR grism covers
four bands, called $SZ$, $J$, $SH$, and $SK$ (referred to as $Z$, $J$, $H$ and
$K$ in this paper). Depending on the order selection filter chosen, this
results in spectral resolutions of 550, 500, 500, and 450, respectively, in
combination with the 1\arcsec\ slit that we have employed (for the $Z$-- and
$J$--band observations of J1030+0524 and J1306+0356, respectively, we used a
narrower 0\farcs6 slit resulting in a slightly higher resolution).  For each
QSO, the \ion{C}{4} and \ion{Mg}{2} lines were observed. Depending on the
redshift, these lines are in the $Z$-- or $J$--band (for J1030+0524), and
$K$--band, respectively.  For J0836+0054, a spectrum in the $H$--band was also
obtained.  Table~\ref{tbl:objects} summarizes the bands chosen and the
exposure times for each objects.  The observations were carried out in service
mode and were collected over several years: some of the data were taken as
early as May 2001, while the bulk was obtained between October 2004 and March
2006.  Exposure times were typically 2.0 hours in $Z$--band and 6.0 hours in
$K$--band, but vary from object to object (see Table \ref{tbl:objects}).

In most cases, twenty frames of 120 seconds were exposed in one hour (or
observation block), in ABBA sequences with large 20\arcsec\ to 30\arcsec\
offsets (dithering) and small random offsets (jittering to avoid bad pixels,
within a box of 4\arcsec\ to 18\arcsec).  A telluric standard was observed
each night (before or after a quasar was observed) at an airmass typically
within 0.2 of the airmass of the QSO observations.

The data were reduced using ESO's pipeline software and IRAF.  Although ESO
delivers the observations together with pipeline reduced products, we have
rereduced the original frames using the newest version of ESO's pipeline
software (for ISAAC, version 5.4.2) to ensure that all spectra were reduced in
the same way.  In particular, care was taken that the pipeline reduction was
carried out using sky emission lines to determine the wavelength calibration
(this was not always the case in the ESO products).  Given the short
integration times of the telluric stars in the $Z$--band, the imprints of the
sky lines were not strong enough to be used as a wavelength indicator, and we
thus used the wavelength calibration given by arc lamps observed the following
day.  The zeropoints of the wavelength scales of the $Z$--band spectra were
then corrected by measuring the position of a single bright sky line and, in a
few cases, of the hydrogen Paschen lines in the continua of stars
serendipitously observed in the same slit as the target.

Rerunning the pipeline produces wavelength calibrated co--added images of the
individual frames observed in one observation block.  Subsequent reduction was
carried out within IRAF.  Using IRAF's {\verb apall } task, one--dimensional
spectra were extracted from these images.  The extraction width was either 5
pixels (0\farcs75) or the seeing FWHM.  The spectra were then corrected for
telluric extinction using IRAF's {\verb telluric } task (by dividing the
observed QSO spectrum by the observed star spectrum), after removal of
instrinsic spectral absorption features from the observed star.  To get the
correct slope and flux calibration, the spectra were multiplied by an
artificial black body spectrum of the standard star based on its spectral type
and magnitude (in the observed band if available from the literature, but
otherwise in the closest band for which a magnitude was available). Then, the
individual one--dimensional spectra were averaged to form a single spectrum per
quasar per band.  The flux calibration of these spectra was checked by
comparison with broad band magnitudes.  As most of the necessary broad band
magnitudes were not available from observations, we fitted the type 1 quasar
template SED constructed by \citet{ric06c} to the available NIR and MIR
photometry from \citet{jia06} and derived $Z$, $J$ and $K$ flux densities from
the resulting normalized SEDs.  As the final step in the data reduction, we
corrected the absolute flux calibration by factors within the range $0.75 -
2.2$ to match these derived broad band fluxes.

\section{Results}\label{sec:results}

\subsection{Composite spectrum}

\begin{figure*}
\includegraphics[angle=0,width=\textwidth]{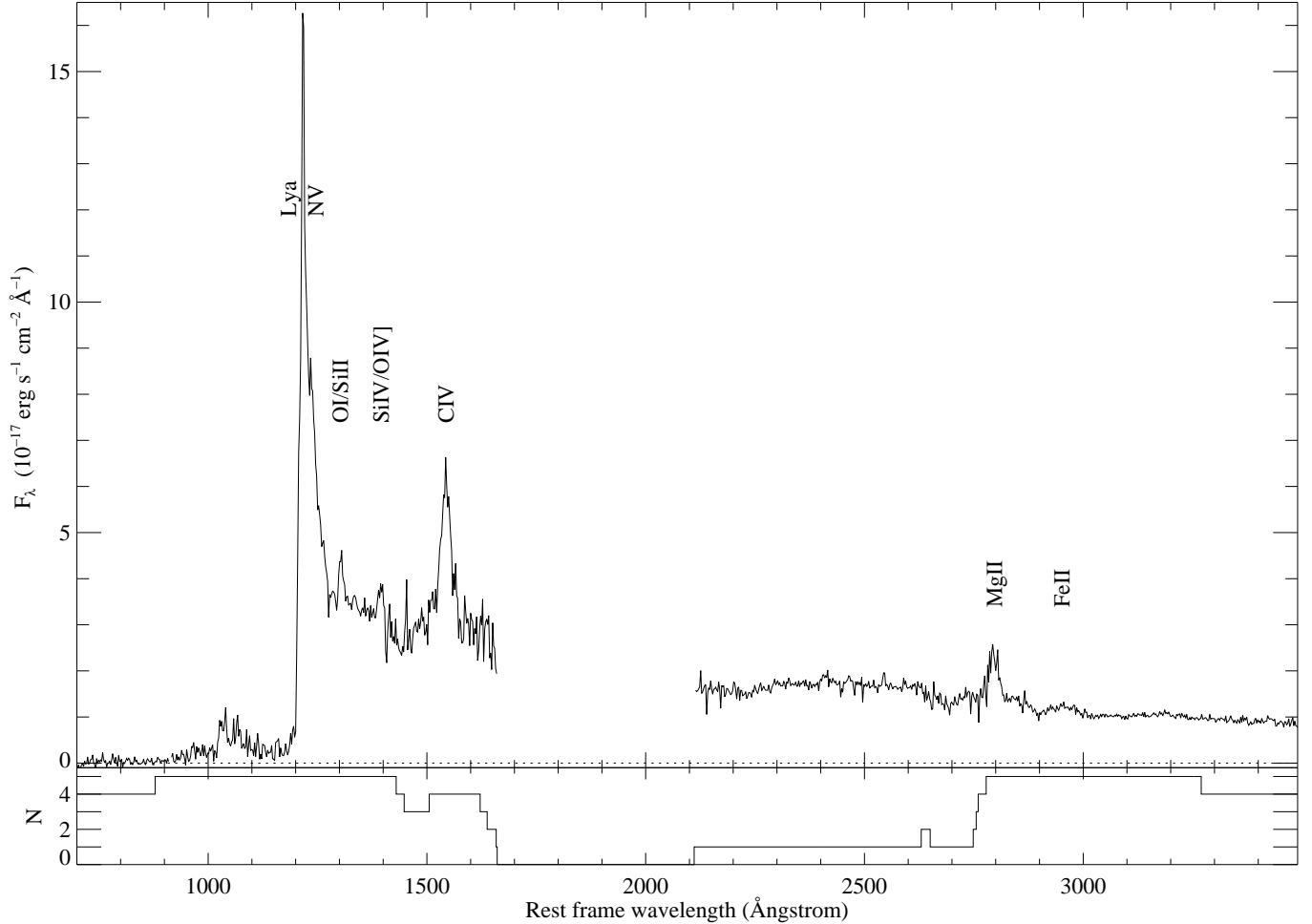}
\caption{Composite spectrum of the five $z \sim6 $ QSOs in our sample at rest
  frame wavelength.  Also displayed is a histogram showing the number of
    spectra contributing at each wavelength.  Strong emission lines are 
  identified.  The depression in flux
  below the \lya\ line shows where the GP trough is located (although this is
  much more pronounced in individual spectra).
  \label{fig:comp_spec}}
\end{figure*}

For illustrative purposes, we have constructed a composite spectrum by
shifting the five QSO spectra to their restframe wavelength and averaging
them, thereby obtaining a spectrum covering a larger wavelength region than
any of the single spectra (Fig.\ \ref{fig:comp_spec}).  The spectra were
normalized to all have the same mean flux density in the restframe wavelength
range $3000 - 3200$\,\AA.  As absolute flux reference in this region was taken
the spectrum of J0836+0054.  Several emission lines are visible, most notable
Ly$\alpha$ and \ion{Si}{4}/\ion{O}{4}] in the observed--frame optical domain,
and \ion{C}{4}, \ion{Mg}{2} and \ion{Fe}{2} in the observed--frame NIR domain.
To measure the properties of these lines, we must determine the underlying
continuum, and, for \ion{Mg}{2}, we also need to determine the Balmer
pseudo--continuum and the \ion{Fe}{2} emission line forest.  This spectral
decomposition is described in the following section.

\subsection{Decomposition of emission lines and continuum}

\subsubsection{Continuum}

The dominant component in the spectra is the continuum, which is usually
modelled as a power law ($F_\lambda \propto \lambda^\alpha$).  For the quasar
composite spectrum obtained from a sample of over 2200 SDSS spectra within a
redshift range $0.044 \leq z \leq 4.789$ (see Fig.\ \ref{fig:sdss_template}),
the slope was found to be $\alpha_\lambda = -1.6$ \citep{vdb01}.  For
individual quasars this value varies, e.g., \citet{ric03} find values between
-1.75 and -1.24 for most of the 4576 SDSS quasars at $0.3 \le z \le 2.2$ in
their sample, \citet{pen03} find a value of -1.43 with a scatter of 0.33 for
45 SDSS quasars at $3.6 \le z \le 5.0$, while \citet{mai04a} find power law
slopes between -1.35 and -2.10 for a sample of eight quasars at $4.9 \leq z
\leq 6.4$.  For spectra with short wavelength coverage and many emission lines
(e.g., \ion{Mg}{2}, \ion{Fe}{2}), the determination of the power law slope is
made difficult by the fact that almost no part of the quasar spectrum is not
also covered by one or several (blended) emission lines.  We, therefore, do
fit a power law to be able to measure the flux in the emission lines
superposed, but note that the measured slope has a large error and might not
be representative for the overall continuum shape of the quasar.

\subsubsection{Balmer pseudo--continuum}

A second component which has to be determined and subtracted in order
to make proper fits to the \ion{Mg}{2} and \ion{Fe}{2} emission
lines, is the forest of hydrogen emission lines, which form the
so--called Balmer pseudo--continuum.

To model the Balmer continuum, we have followed \citet{die03b}, who
assume gas clouds of uniform temperature ($T_e = 15000$ K) that are
partially optically thick.  The Balmer continuum spectrum below the
Balmer edge ($\lambda_{\rm BE} = 3646$ \AA) can be described by
\begin{equation} F{_\lambda}^{\rm BaC} = F^{\rm BE}
  B_\lambda(T_e)(1-e^{-\tau_{\rm BE} (\lambda/\lambda_{\rm BE})^3}), 
  \lambda \leq \lambda_{\rm BE}
\end{equation}
where $B_\lambda(T_e)$ is the Planck function at the electron temperature
$T_e$, $\tau_{\rm BE}$ is the optical depth at $\lambda_{\rm BE}$, and $F^{\rm
  BE}$ is the normalized flux density at the Balmer edge \citep{gra82}.  In
general, the normalization is estimated at $\lambda \simeq 3675$ \AA\ since no
\ion{Fe}{2} emission is present at that wavelength.  However, in our spectra,
this region is either absent or has a low signal--to--noise ratio.  Therefore,
the normalization is fixed to the strength of the power--law continuum during
the fit and the optical depth is fixed at $\tau_{\rm BE} = 1$ (we found
minimal differences in the Balmer continuum when we varyied $\tau_{\rm BE}$
between 0.1 and 2.0).

\subsubsection{\ion{Fe}{2} template}

The \ion{Fe}{2} ion emits a forest of lines, many of which are
blended.  These are usually fit jointly by using
an iron emission line template.  Several templates are available in
the literature, based on high signal--to--noise ratio spectra of individual
active galaxies \citep[e.g.,][]{ves01}, on composites of many quasar
spectra \citep[e.g., from the LBQS by \citealt{fra91} and from the
SDSS by][]{vdb01}, or on theoretical models \citep[e.g.,][]{ver99,sig03}.
Before fitting our data, we have compared two templates applicable to
high redshift quasar spectra: one derived from the nearby narrow line
Seyfert 1 galaxy, I Zwicky 1, published by \citet{ves01} and one we
derived from the composite SDSS QSO spectrum \citep{vdb01}.

\vspace{1ex}
\centerline{\emph{I Zw 1 template by \citeauthor{ves01}}}
\vspace{1ex}

\citeauthor{ves01} have constructed an \ion{Fe}{2} and \ion{Fe}{3} template
based on \emph{Hubble Space Telescope} spectra of the narrow line Seyfert 1
galaxy, I Zwicky 1 (I Zw 1, PG 0050+124, $z = 0.061$).  Its narrow intrinsic
lines ($\sim 900$ km s$^{-1}$) and its rich iron spectrum make the template
particularly suitable for fitting the spectra of active galactic nuclei and
quasars.  \citeauthor{ves01} fitted and subtracted a power law continuum, and
absorption and emission features from other elements than iron from the I Zw 1
spectrum.  An \ion{Fe}{3} emission model was subtracted from the residual to
create a pure \ion{Fe}{2} template.  In the process, the \ion{Mg}{2}
$\lambda\lambda$2795,2803 emission was fit with two double Gaussian components
400 km s$^{-1}$ apart. After subtraction of this \ion{Mg}{2} fit, no iron
emission is left in the $2770-2820$\,\AA\ range, although from theoretical
considerations, it should not be entirely absent.  Thus, subtraction of their
Fe template would leave some unidentified iron emission around 2800\,\AA,
leading to an overestimate of the flux of the \ion{Mg}{2} line.  We have
therefore added a constant flux density to the \citeauthor{ves01} template of
$\sim$20\% of the mean flux density of the template between 2930 and 2970 \AA,
as illustrated in Fig.~\ref{fig:vestergaard_sdss_comparison} \citep[roughly
consistent with the value envisaged by][]{ves01}.  The justification for this
operation comes from the theoretical \ion{Fe}{2} emission line strength by
\citet[][see their Fig.~13]{sig03}.

\vspace{1ex}
\centerline{\emph{SDSS QSO composite template by \citeauthor{vdb01}}}
\vspace{1ex}

\begin{figure}
\includegraphics[angle=90,width=\columnwidth]{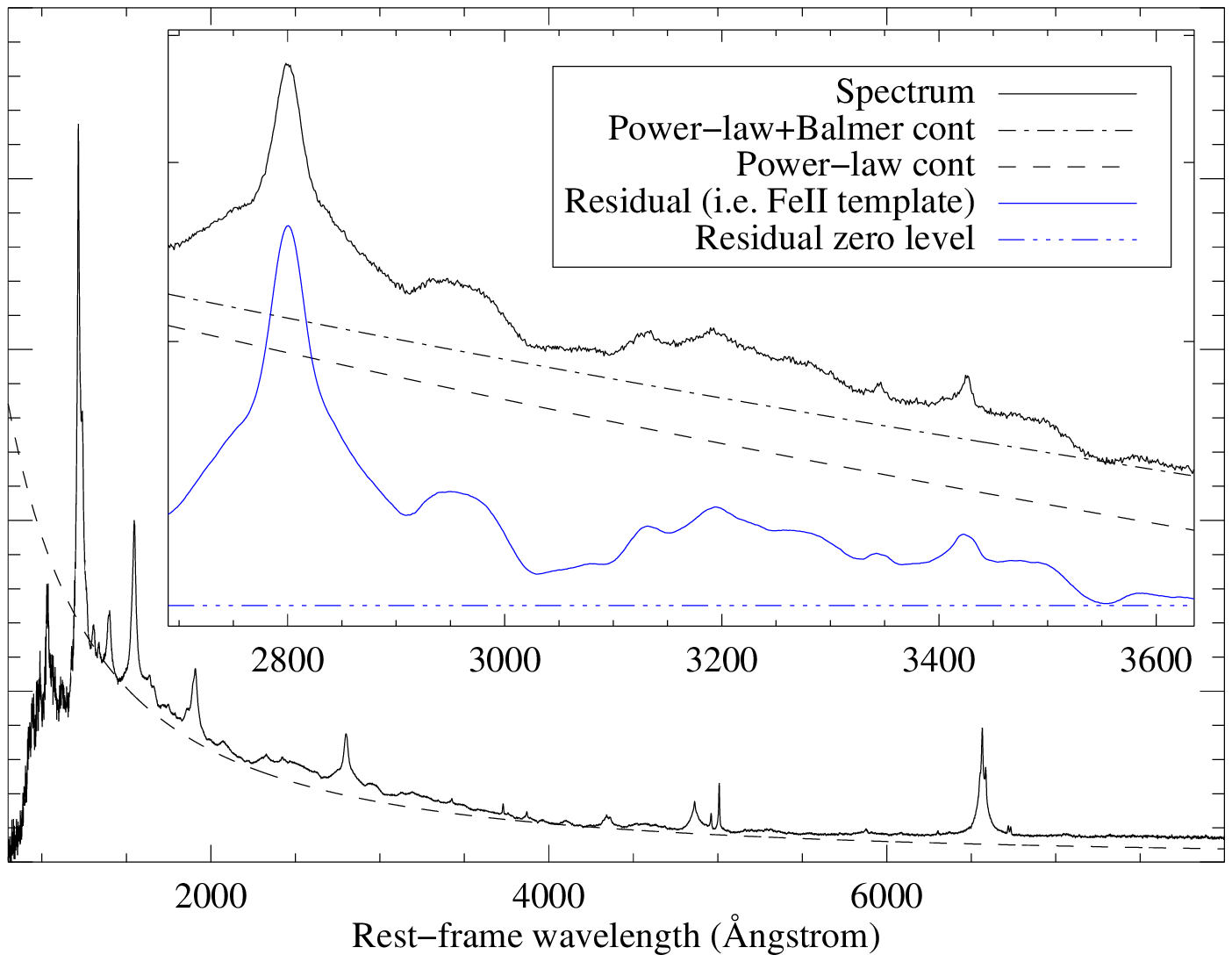}
\caption{The SDSS median quasar composite spectrum \citep{vdb01}, shown 
    with an arbitrary flux scaling.  The
    dashed line is a power law continuum ($F_\lambda =
    \lambda^\alpha$) with a slope $\alpha = -1.54$ given by
    \citeauthor{vdb01}\ and a normalization fit by us.  The inset shows
    a zoom in the region of the \mgt\ and part of the
    \fet\ emission.  Also shown is the power law continuum and
    the co--added power law and Balmer continua, which are subtracted
    from the smoothed composite spectrum to form the final \fet\
    emission template, shown in grey [blue].  [See the electronic 
    edition of the Journal for a color version of this figure.]
    \label{fig:sdss_template}}
\end{figure}

\begin{figure}
\includegraphics[angle=90,width=\columnwidth]{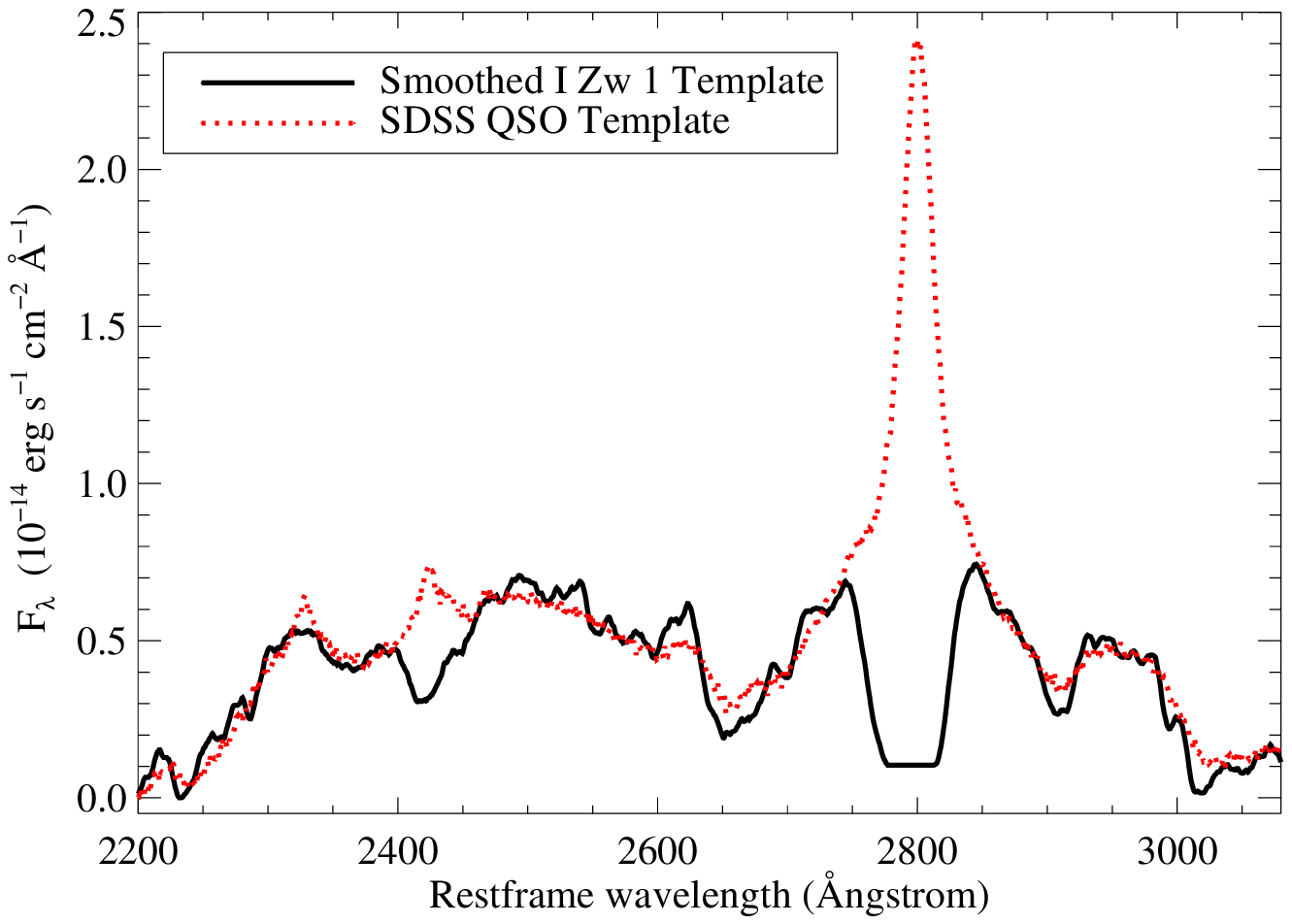}
\caption{Our modified \fet\ \citet{ves01} template (solid line, smoothed here
  for display purposes), compared with our template derived from the SDSS
  composite QSO spectrum (indicated by the dashed [red] line,
  \citealt{vdb01}).  The templates differ significantly only around 2418\,\AA,
  where the Fe\,{\scriptsize III}\,UV47 line is located and around 2800\,\AA,
  where the prominent \mgt\ line is located.  These lines are not present in
  the \citeauthor{ves01} template which contains only \fet\ emission.  [See
  the electronic edition of the Journal for a color version of this figure.]
  \label{fig:vestergaard_sdss_comparison}}
\end{figure}

From the composite SDSS spectrum, smoothed to a resolution of 7\,\AA,
comparable to the observed rest frame resolution, we have subtracted a power
law continuum with the slope given by \citet{vdb01}.  We have normalized the
power law to the measured values in the $4200-4230$\,\AA\ region \citep[as
specified in][]{vdb01}.  In addition, a Balmer continuum was created and
subtracted, normalized to the measured value at 3675 \AA\ and with optical
depth $\tau_{\rm BE} = 1.0$ (again, only minimal differences were found when
changing $\tau_{\rm BE}$ within reasonable limits).  The result is shown as an
inset in Fig.~\ref{fig:sdss_template}.

\begin{center}
\emph{Comparison of the I Zw 1 and SDSS QSO composite\\
 templates}
\end{center}

We have subsequently compared the \citeauthor{ves01} and the SDSS
templates (see Fig.\ \ref{fig:vestergaard_sdss_comparison}).  The figure
shows that the two templates are very similar, apart from the
\ion{Mg}{2} line which we did not try to subtract in the SDSS template
and an \ion{Fe}{3}\,UV47 line at $\lambda = 2418$ \AA.  We have therefore
decided to use the \citet{ves01} template for the analysis
that follows, as it allows us to fit the \ion{Mg}{2} line simultaneously.

\subsection{Fitting procedure}

Fitting was carried out with the \verb=MPFIT=\footnote{MPFIT is an IDL routine
  written and distributed by C.~Markwardt
  (\url{http://cow.physics.wisc.edu/$\sim$craigm}).} package which performs the
Levenberg--Marquardt least--squares minimization.  Nominal errors of the fitted
parameters were computed by multiplication of the formal one--sigma errors
computed from the covariance matrix with the square root of the reduced
$\chi^2$,  which was between 1.0 and 1.5 in all cases.

\begin{figure*}
\includegraphics[angle=0,width=\textwidth]{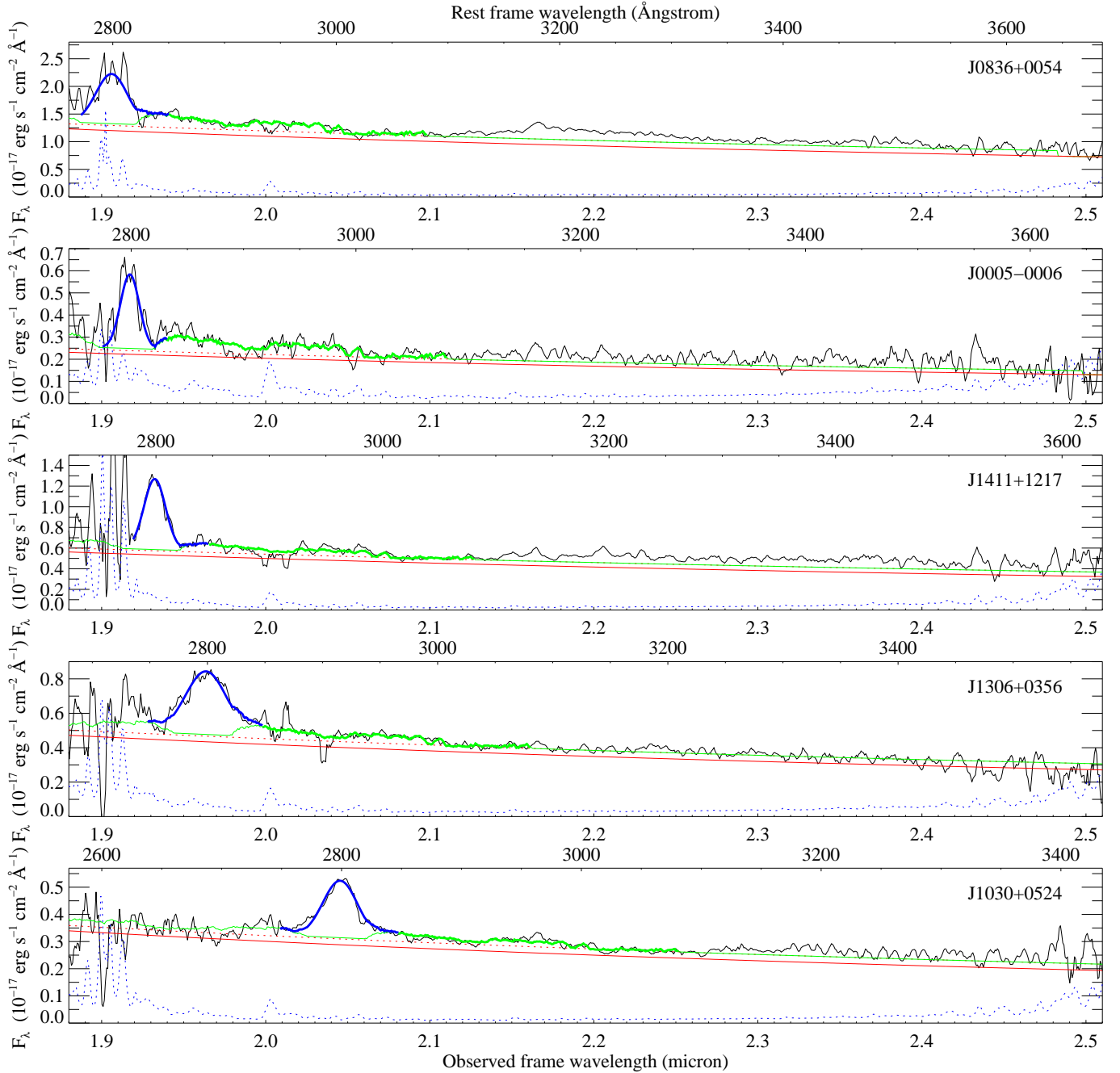}
\caption{$K$--band spectra of the five $z \sim 6$ QSOs, covering the
  wavelength region that includes \mgt, shown here smoothed over 5
  pixels (or 0.0036\,$\mu$m).  The observed wavelength is indicated below the
  spectra (in microns), while the rest frame wavelength is indicated above the
  spectra (in \AA).  Several features of the fit are indicated on top of the
  spectra (dark [black] solid line): the power law continuum (grey [red] solid
  line), the Balmer pseudo--continuum (grey [red] dashed line), Vestergaard
  \fet\ template (light [green] solid line, thick where it is actually
  fitted to the spectrum), the Gaussian fit to the \mgt\ line (dark
  [blue] thick line).  The dashed grey [blue] line shows the noise level per
  pixel (smoothing is not taken into account here).  [See the electronic
  edition of the Journal for a color version of this figure.]
  \label{fig:mgii_fits}}
\end{figure*}

The $K$--band spectra were fitted in two steps (see Fig.~\ref{fig:mgii_fits}).
First, the part of the spectra corresponding to rest--frame wavelengths in the
range 2848 to 3081\,\AA\ was fit simultaneously by the power--law continuum,
Balmer pseudo--continuum and iron template.  Here, the Balmer continuum
normalization was fixed to the power--law continuum strength at 3600\,\AA,
which leaves only three free parameters being fit at this stage: the slope and
normalization of the power--law continuum, and the normalization of the iron
template.  The iron normalization is mostly determined by the emission feature
at $\sim2950$\,\AA.  We did not attempt to fit the redshift of the iron
template.  Instead it was kept fixed (to the \ion{Mg}{2} redshift) and only
the template normalization was varied (a few iterations were necessary to
obtain this redshift).  The fit was then extrapolated to the full spectral
range of the $K$--band spectrum and subtracted from the data.  The residual
still contains the \ion{Mg}{2} line which was subsequently fit with a Gaussian
function, having three free parameters: central wavelength, width and
normalization.  The \ion{Fe}{2} flux was computed by multiplying the derived
iron template normalization with the total \ion{Fe}{2} flux derived from the
iron template by integrating it over the wavelength range $2200 < \lambda <
3090$\,\AA.  Here, the assumption is made that the model (the iron template)
represents the observed data well.  The $\chi^2$ obtained during the fitting
procedure gives an indication for this representativeness, but only for the
relatively small range $2848 < \lambda < 3081$\,\AA.  \citet{ves01} state that
the iron template derived from I Zw 1 is generally applicable to high redshift
quasar spectra.  Our comparison to the SDSS template confirms this.  To
account for this uncertainty, however, we multiply the nominal error in the
\ion{Fe}{2} flux by a factor two to obtain the error displayed in
Table~\ref{tbl:mgii_fits}.

\begin{figure*}
\includegraphics[angle=0,width=\textwidth]{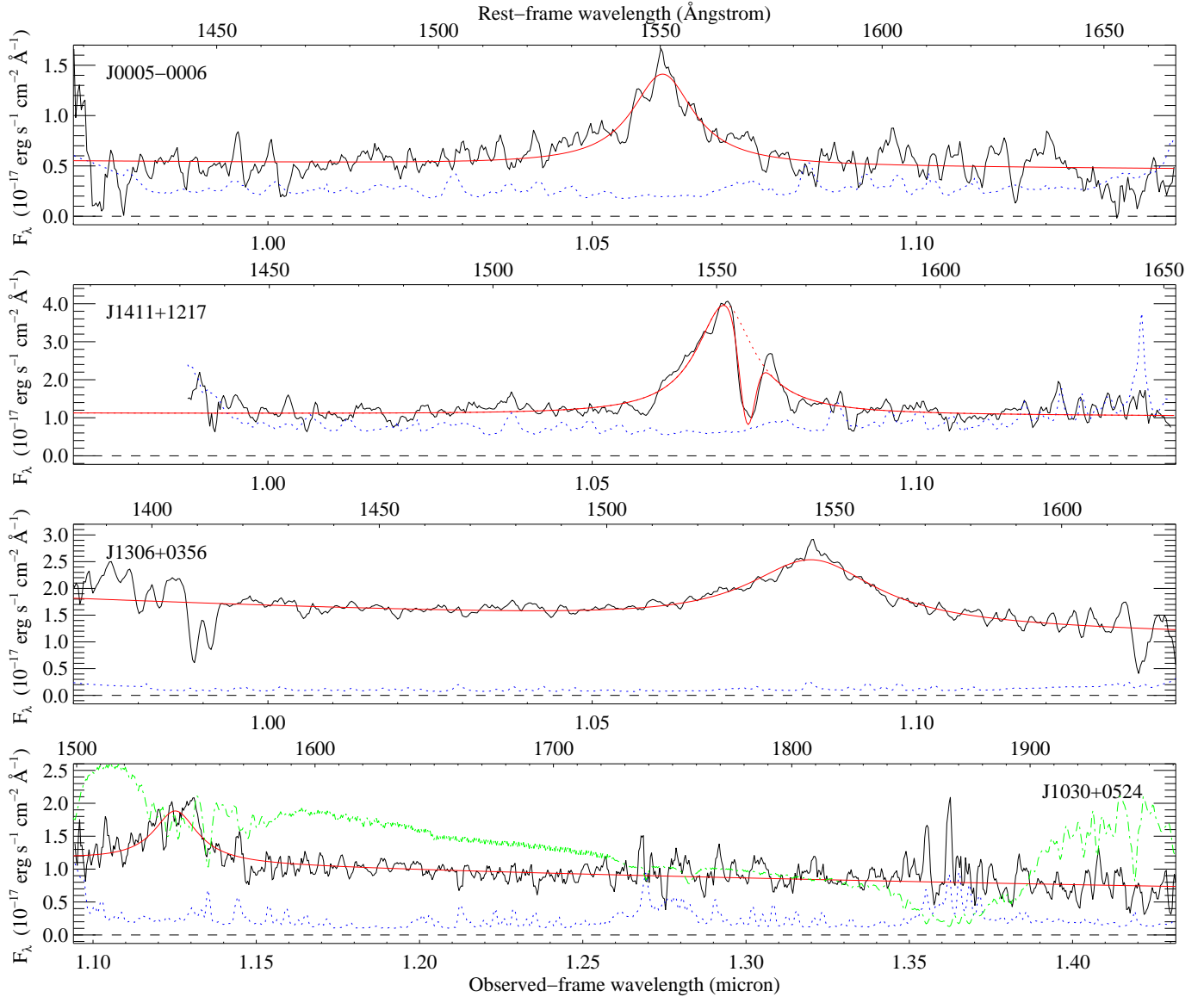}
\caption{$Z$-- and $J$--band spectra of the five $z \sim 6$ QSOs, 
  covering the wavelength region that includes \ct, smoothed over 
  five pixels (or 0.0014\,$\mu$m).  Overplotted is the simultaneous fit 
  of a power law and
  Lorentzian curve (solid [red] line).  The absorption feature within the
  \ct\ line in J1411+1217 was fitted simultaneously with a Voigt
  function.  The unabsorbed Lorentz profile for J1411+1217 is shown with a
  dashed line.  The dashed [blue] line shows the noise level per pixel
  (smoothing is not taken into account here).  Overplotted on the spectrum of
  J1030+0524 in grey [green] is the telluric transmission measured, showing
  the problematic telluric absorption at 1.136\,$\mu$m. [See the electronic
  edition of the Journal for a color version of this figure.]
  \label{fig:civ_plots}}
\end{figure*}

The $Z$ and $J$--band spectra were fitted in one step (see
Fig.~\ref{fig:civ_plots}).  The \ion{C}{4} line was fit with a Lorentzian
function, while the underlying continuum was fit with a power law.  The
function fit has therefore five free parameters: the power law slope and
normalization, and the Lorentzian function's normalization, width and central
wavelength.  J1411+1217 shows associated absorption in the \ion{C}{4} line,
which was fit with a Voigt profile, adding four more free parameters.  The
fitting parameters are summarized in Table~\ref{tbl:civ_fits}.

\begin{deluxetable*}{lrrrrrrrr }
\tabletypesize{\scriptsize}
\tablecaption{Fitted spectral properties of the Fe\,{\scriptsize II} and Mg\,{\scriptsize II} lines and Fe\,{\scriptsize II}/Mg\,{\scriptsize II} ratio\label{tbl:mgii_fits}}
\tablewidth{0pt}
\tablehead{
\colhead{QSO} & \colhead{$z$\tablenotemark{a}} & 
\colhead{$\alpha$\tablenotemark{b}} & 
\colhead{$F_{\rm FeII}$\tablenotemark{c}} & 
\colhead{$z_{\rm MgII}$\tablenotemark{d}} & 
\colhead{$F_{\rm MgII}$\tablenotemark{e}} & 
\colhead{EW$_{\rm 0,MgII}$\tablenotemark{f}} & 
\colhead{FWHM$_{\rm MgII}$} & 
\colhead{Fe\,{\scriptsize II}/Mg\,{\scriptsize II}} \\
&&& \colhead{10$^{-15}$} && \colhead{10$^{-15}$} \\
&&
\multicolumn{2}{r}{[erg\,cm$^{-2}$\,s$^{-1}$]} & 
\multicolumn{2}{r}{[erg\,cm$^{-2}$\,s$^{-1}$]} &
\colhead{[\AA]} & \colhead{[km\,s$^{-1}$]}
}
\startdata
J0836+0054 & 5.82 & -1.81 & 7.4$\pm$1.2 & 5.810$\pm$0.003 & 2.15$\pm$0.17 & 24$\pm$3 & 3600$\pm$300 & 3.4$\pm$0.6 \\
J0005$-$0006 & 5.85 & -2.05 & 2.3$\pm$0.6 & 5.850$\pm$0.003 & 0.48$\pm$0.06 & 29$\pm$5 & 2100$\pm$300 & 4.7$\pm$1.5 \\
J1411+1217 & 5.95 & -1.96 & 2.7$\pm$0.7 & 5.904$\pm$0.002 & 1.13$\pm$0.07 & 28$\pm$3 & 2400$\pm$150 & 2.4$\pm$0.6 \\
J1306+0356 & 5.99 & -1.94 & 2.4$\pm$0.9 & 6.016$\pm$0.002 & 1.14$\pm$0.04 & 35$\pm$2 & 4500$\pm$160 & 2.1$\pm$0.8\\
J1030+0524 & 6.28 & -1.94 & 1.2$\pm$0.3 & 6.308$\pm$0.001 & 0.55$\pm$0.01 & 25$\pm$1 & 3600$\pm$100 & 2.1$\pm$0.5
\enddata
\tablenotetext{a}{The redshift as published in the discovery paper, based on the observed wavelength of the \lya\ line.} 
\tablenotetext{b}{Slope of the power law continuum: $F_\lambda \propto 
\lambda^\alpha$.}
\tablenotetext{c}{Flux derived from the Fe\,II template normalization
(derived over the range $2850 \lesssim \lambda_0 \lesssim 3070$ \AA).}
\tablenotetext{d}{The redshift derived from the Mg\,II centroid.}
\tablenotetext{e}{Flux derived from the Mg\,II line fit.} 
\tablenotetext{f}{Rest frame equivalent width of Mg\,II line in \AA.}
\end{deluxetable*}

\begin{deluxetable}{lrrrrr}
\tabletypesize{\scriptsize}
\tablecaption{Fitted spectral properties of the C\,{\scriptsize IV} line\label{tbl:civ_fits}}
\tablewidth{0pt}
\tablehead{
\colhead{QSO} & \colhead{$z$\tablenotemark{a}} & 
\colhead{$z_{\rm CIV}$\tablenotemark{b}} & 
\colhead{$F$\tablenotemark{c}} & 
\colhead{EW$_{\rm 0}$\tablenotemark{d}} &
\colhead{FWHM} \\
&&& \colhead{10$^{-15}$} \\
&&
\multicolumn{2}{r}{[erg\,cm$^{-2}$\,s$^{-1}$]} & 
\colhead{[\AA]} & \colhead{[km s$^{-1}$]}
}
\startdata
J0005$-$0006 & 5.85 & 5.848$\pm$0.001 & 1.5$\pm$0.1 & 40$\pm$2 & 2900$\pm$140 \\
J1411+1217   & 5.95 & 5.911$\pm$0.001 & 5.0$\pm$0.1\tablenotemark{e}
\hspace{-2.1ex}                   & 70$\pm$3 & 3060$\pm$ 80 \\
J1306+0356   & 5.99 & 5.997$\pm$0.001 & 3.9$\pm$0.2& 40$\pm$2 & 5940$\pm$200 \\
J1030+0524   & 6.28 & 6.262$\pm$0.003 & 2.4$\pm$0.2& 32$\pm$3 & 5030$\pm$380
\enddata
\tablenotetext{a}{The redshift as published in the discovery paper.}  
\tablenotetext{b}{The redshift derived from the centroid of the 
C\,IV line.}
\tablenotetext{c}{Flux derived from the C\,IV line fit.} 
\tablenotetext{d}{Rest frame equivalent width in \AA.}
\tablenotetext{e}{Unabsorbed value, actual observed value is smaller.}
\end{deluxetable}

\section{Discussion}\label{sec:discussion}
\subsection{Metal enrichment}

As described in Sec.\ \ref{sec:introduction}, a measurement of the Fe to Mg
ratio at $z \sim 6$ may serve as an approximate indication of the onset of
star formation in the highest redshift quasars.  The measured
\ion{Fe}{2}/\ion{Mg}{2} line ratios of our five $z \sim 6$ QSOs are shown in
Table \ref{tbl:mgii_fits}.  The ratios lie within the range $2 - 5$.

\begin{figure}
\includegraphics[angle=00,width=\columnwidth]{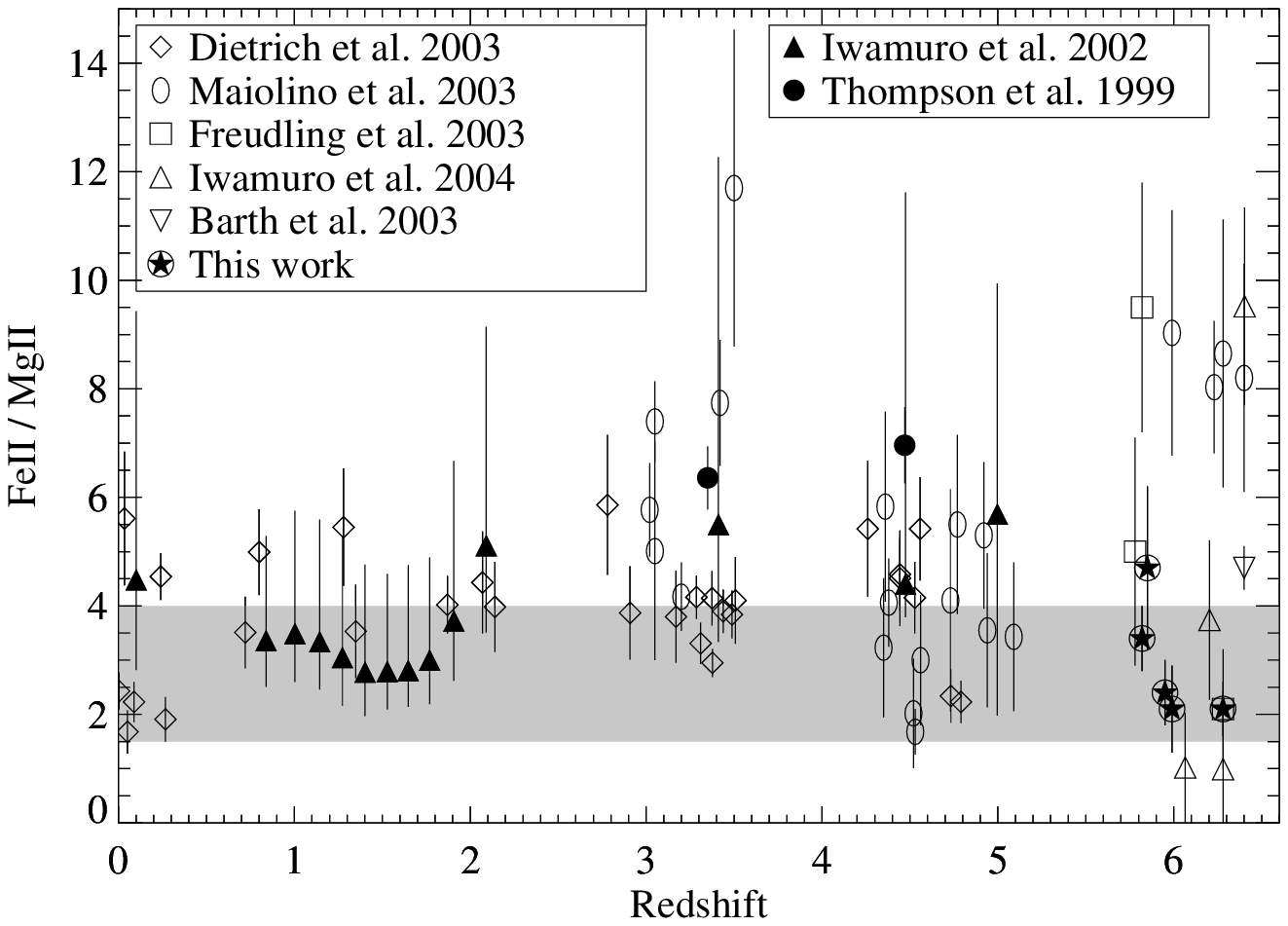}
\caption{\fet/\mgt\ line ratio plotted versus redshift
  of our sample of quasars (encircled stars) compared with quasars
  at redshifts $0 < z < 6.4$ from the literature: \citet[][diamonds,
  including \citealp{die02}]{die03b}, \citet[][squares]{fre03},
  \citet[][triangles]{iwa04}, \citet[][circles]{tho99},
  \citet[][upside down triangle]{bar03}, \citet[][ellipses]{mai03}.
  Filled symbols represent sample averages and their error bars
  represent the standard deviation of the averages of the measured
  values.  The range of values obtained by modelling the \fet\ and
  \mgt\ line strengths of gas with cosmic abundance and a range of
  ionizing continua \citep[2.75$\pm$1.25,
  see][]{wil85} is indicated by a grey band. See text for details
  regarding the difference in the determination of this ratio employed
  by various authors.
  \label{fig:femg_plot}}
\end{figure}

We have compared the \ion{Fe}{2}/\ion{Mg}{2} line ratio measured in our study
to those of other studies at similar and lower redshifts (summarized in
Fig.~\ref{fig:femg_plot}).  It is, however, difficult to make a direct
comparison: different studies use different wavelength regions to estimate the
underlying continuum and \ion{Fe}{2} features and use different methods to fit
the spectral features.  \citet{iwa02} try to overcome this problem by applying
the same fitting algorithm to new spectra at $4.4 < z < 5.3$ and archival
spectra at lower redshifts down to $z = 0.03$.  Although the scatter within
their redshift bins is large, they measure a median value of
\ion{Fe}{2}/\ion{Mg}{2} at $z \sim 5$ which is about 50\% higher than at $z
\sim 1.5$.  Several other authors have studied the \ion{Fe}{2}/\ion{Mg}{2}
line ratio for QSOs at higher redshifts: $5.7 < z < 6.3$ \citep{fre03}, $3.0 <
z < 6.4$ \citep{mai03} and $6.1 < z < 6.3$ \citep{iwa04}.  The latter paper
summarizes these data and concludes that the upper envelope of the
\ion{Fe}{2}/\ion{Mg}{2} distribution decreases towards higher redshift ($z >
3$) while the scatter among individual objects increases.  This may indicate
that some objects are observed such a short time after the initial starburst
that the Broad Line Region (BLR) is not yet fully enriched with iron.
Fig.~\ref{fig:femg_plot} shows that our measurements confirm the lack of
evolution in the \ion{Fe}{2}/\ion{Mg}{2} line ratio observed by other authors,
indicating early enrichment of the BLR.  As the age of the Universe at $z =
6.0$ is only 0.9 Gyr, an enrichment time of 1 Gyr needed to reach the observed
ratio according to some models \citep[e.g.,][]{yos98} seems to be ruled out.
Assuming an enrichment time of 0.3 Gyr appropriate for an elliptical galaxy
\citep{mat01}, the onset of star formation would have been at $z \gtrsim 8$.

\citet{wil85} discuss how well the \ion{Fe}{2}/\ion{Mg}{2} line ratio
represents the Fe/Mg abundance ratio, as the region where \ion{Fe}{2} is
produced is more extended than the \ion{Mg}{2} region, and the radiative
transfer of the lines is very different.  Computing the \ion{Fe}{2} and
\ion{Mg}{2} line strengths for a range of hydrogen densities and ionization
parameters and for a cosmic abundance of elements, they predict values between
1.5 and 4 for the line ratio and attribute higher ratios to an overabundance
of iron, with respect to magnesium, but probably also with respect to
hydrogen.  These computations are, however, based on a limited 70--level model
of the iron atom.  \citet{bal04} construct a 371--level Fe$^+$ model, using
energy balance to obtain a self--consistent temperature and ionization
structure.  They apply this model to a large range of parameters in order to
reproduce the observed \ion{Fe}{2} emission properties and the relative
strengths of the strong emission line of other elements (such as Ly$\alpha$
and \ion{Mg}{2}).  Only two scenarios for observed quasar spectra are allowed
by the results of this modeling: there is either significant microturbulence
($v_{\rm turb} \ge 100$ km\,s$^{-1}$), consistent with currently popular
models where the emission lines are formed in a wind flowing off a rotating
accretion disk, or the \ion{Fe}{2} lines are formed in gas with other
properties than the gas where the other emission lines are formed, i.e., in
spatially separated regions.  The bottom line is that, although the relative
strength of \ion{Fe}{2} does depend somewhat \citep[but not
linearly,][]{ver99} on the iron abundance, it also depends sensitively on
other parameters, preventing strong conclusions on the Fe/Mg abundance ratio
from the \ion{Fe}{2}/\ion{Mg}{2} line ratio, including the comparison with the
presumed \emph{solar} value (see Fig.~\ref{fig:femg_plot}).

Although our \ion{Fe}{2}/\ion{Mg}{2} line ratios are consistent with the
values derived by \citet{iwa04} and \citet{bar03}, they are systematically
lower than those measured by \citet{mai03}.  For J1030+0524 our ratio is
consistent with that found by \citet{fre03}, while for J0836+0054 it is lower.
Note that, using the original \ion{Fe}{2} template provided by \citet{ves01}
would result in even lower \ion{Fe}{2}/\ion{Mg}{2} ratios for our sample (as
we would have overestimated the contribution of the \ion{Mg}{2} line
emission).  Within individual studies, the emission line ratios for different
$z \sim 6$ quasars also span a large range of values (e.g., 1.0 to 9.5 for
\citeauthor{iwa04}\ and 2.1 to 9.5 for \citeauthor{fre03}), indicating that
there is significant scatter within the population of $z \sim 6$ quasars.
There are several possible causes for the observed discrepancies between
different authors, including the iron template employed, the wavelength limits
over which the template is integrated, the wavelength range over which the
template is actually fit, and the length and signal--to--noise ratio of the
spectra used.  To derive clear conclusions on the evolution of the
\ion{Fe}{2}/\ion{Mg}{2} line ratio, a better comparison study should be made
where all spectra available in the literature are fit by the same template
using the same method, but this work is beyond the scope of this paper.

We have found no strong correlations between \ion{Fe}{2}/\ion{Mg}{2} line
ratio, continuum power law slope $\alpha$ and luminosity within our sample.
The scatter in the \ion{Fe}{2}/\ion{Mg}{2} line ratio seems to be small in our
sample and independent of luminosity, although the lowest luminosity object is
also the iron richest.  However, this object, J0005$-$0006, seems to be
peculiar, as it has narrower UV emission lines \citep[$1500 -
2500$\,km\,s$^{-1}$,][]{fan04} than other SDSS QSOs and may not be
representative of the general $z \sim 6$ QSO population.

We here summarize the (in)consistencies in measurements by different
authors for individual objects in our sample.

\vspace{1ex}
\centerline{\emph{Notes on individual objects}}
\vspace{1ex}

\noindent
{\bf J0836+0054} This QSO, at $z = 5.82$ was observed in the NIR by
\citet{fre03} and \citet{ste03}.  \citet{fre03} find a (scaled)
\ion{Fe}{2}/\ion{Mg}{2} line ratio of 9.5$\pm$2.3, while the spectrum of
\citet{ste03} does not include the $K$--band.  We find a ratio of 3.4$\pm$0.6,
significantly lower than that found by \citeauthor{fre03}\smallskip

\noindent
{\bf J1306+0356} This QSO, at $z = 5.99$ was observed in the NIR by
\citet{mai03}, who found a ratio of 9.0$\pm$2.3, much higher than
the value of 2.1$\pm$0.8 found by us.\smallskip

\noindent {\bf J1030+0524} The most distant QSO of our sample, at $z = 6.28$
was observed in the NIR by \citet{mai03} and \citet{pen02}.  \citet{mai03},
again, find a much higher ratio of 8.7$\pm$2.5 than ours of 2.1$\pm$0.5,
which, however, is consistent with the value of 2.1$\pm$1.1 derived by
\citet{fre03} from a NICMOS spectrum sampling rest--frame wavelengths between
1600 and 2700\,\AA.  The ratio derived by us is also consistent with the ratio
of $0.99^{+1.86}_{-0.99}$ reported by \citet{iwa04}.

\subsection{Black hole masses}

The enormous energy output of QSOs is most probably powered by a central black
hole which is surrounded by an accretion disc.  The mass of this black hole is
a fundamental parameter which, at least in nearby galaxies, is strongly
correlated with the bulge velocity dispersion of the galaxy host \citep[the so
called $M_{\rm BH}-\sigma_{\rm bulge}$ relation, where $\sigma_{\rm bulge}$ is
the velocity dispersion of the host galaxy's bulge,
e.g.,][]{geb00,fer00,tre02,woo06}.  This relation suggests that the evolution
of the black hole and the build--up of the bulge in its host galaxy are
intimately related.  The central black hole mass of our Galaxy was measured by
very high spatial resolution imaging and subsequent modelling of the stellar
dynamics \citep{sch03}, but this method cannot be applied to extragalactic
black holes as the required spatial resolution cannot be attained.  The most
direct measurements of central black holes mass in AGN come from reverbaration
mapping studies \citep[e.g.,][]{pet04,kas07}, but these time intensive
observations are not easily feasible for very distant QSOs.  Here, we derive
black hole masses using three different approaches, as described below.

\subsubsection{Eddington mass}
The simplest method to estimate the black hole mass is to use the bolometric
luminosity of the QSO and to assume that the outward radiation pressure
gradient acting on the gas is just enough to counterbalance the gravitational
attraction of the black hole.  This gives an estimate of the black hole mass,
called the Eddington mass \citep{edd22,mar73}
\begin{equation}
L_{\rm Edd} = 1.3 \times 10^{38} 
{\left(M_{\rm BH} \over M_\odot\right)}\, {\rm erg\,s^{-1}},
\end{equation}
i.e., the black hole mass $M_{\rm BH}$ can be derived assuming $L_{\rm
bol} = L_{\rm Edd}$.

\subsubsection{Black hole mass derived from \ion{Mg}{2}}
Under the assumption that the dynamics of the broad line region are
dominated by the gravity of the central black hole, an estimate of the black
hole mass is given by $M_{\rm BH} \approx G^{-1} R_{\rm BLR} V^2$, where
$M_{\rm BH}$ is the black hole mass, $R_{\rm BLR}$ the radius of the BLR and
$V$ the velocity of the gas, which can be estimated from the width of emission
lines in the QSO's spectrum (see \citealt{mcl04} and references therein for a
discussion of this assumption and of the emission lines which can be used).
The radius of the BLR is strongly correlated with the AGN monochromatic
continuum luminosity at 3000 \AA\ \citep{mcl02}, allowing black hole mass
estimates for high redshift quasars to be made from a single spectrum covering
the blended \ion{Mg}{2}$\lambda$\,2795,2803 line doublet and redward
continuum.

\citet{mcl04} use 17 AGN (all at $z < 0.7$) with luminosities
comparable to those of known high redshift QSOs ($\lambda$L$_\lambda >
10^{37}$ W) for which reverberation mapping estimates of the BH mass
are available \citep[from][and \citealp{kas00}]{wan99}, to determine
the following relation between black hole mass, continuum luminosity
at $\lambda_0 = 3000$ \AA\ and \ion{Mg}{2} line width:
\begin{equation}
{M_{\rm BH} \over {\rm M_\odot}} = 3.2\,
{\left\lgroup\lambda L_{3000} \over 10^{37}\,{\rm W}\right\rgroup}^{0.62} 
{\left\lgroup {\rm FWHM\,[MgII]} \over {\rm km\,s^{-1}}\right\rgroup}^2,
\end{equation}
where $L_{3000}$ is the luminosity per wavelength bin at rest--frame
$\lambda_0 = 3000$ \AA\ and FWHM is the Full Width at Half Maximum of the
\ion{Mg}{2} line.  \citet{mcl02} state that an earlier version of this black
hole mass estimator can reproduce the reverberation black hole mass to within
a factor 2.5 ($1\sigma$).

\subsubsection{Black hole mass derived from \ion{C}{4}}
\citet{ves06} use 32 AGN with reliable reverberation mapping estimates
calculated by \citet{pet04} to derive the following empirical relation
between black hole mass, continuum luminosity at $\lambda_0 = 1350$
\AA, and \ion{C}{4} line width:
\begin{equation}
{M_{\rm BH} \over {\rm M_\odot}} = 4.6\,
{\left\lgroup\lambda L_{1350} \over 10^{37}\,{\rm W}\right\rgroup}^{0.53} 
{\left\lgroup {\rm FWHM\,[CIV]} \over {\rm km\,s^{-1}}\right\rgroup}^2.
\end{equation}
\citeauthor{ves06} estimate that this mass estimator is accurate within
a factor $3.6-4.6$, but they also note that the reverberation--based masses
themselves are uncertain typically by a factor $\sim 2.9$.

It should be noted that some authors \citep[e.g.,][]{bas05} have argued that
the \ion{C}{4} originates, at least in part, from outflowing gas or that the
shift and asymmetry commonly seen in \ion{C}{4} suggest that non--gravitational
effects, such as obscuration and radiation pressure, may affect the line
profile \citep[e.g.,][]{ric02b} and its FWHM would therefore not be a good
measure of the gas dynamics in the BLR, although \citet{ves06} claim that the
typical uncertainties resulting from these are smaller than those quoted above
for the mass estimator itself.

\subsubsection{Derived black hole masses}
Using these three approaches, we estimate black hole masses for our sample and
list them in Table \ref{tbl:bh_masses}.  The bolometric luminosities for these
quasars were obtained by \citet{jia06} from SED template fits to combined
observations at wavelengths including the radio, mm/submm, UV/optical, NIR,
MIR and X--rays.  For the computation of $L_{1350}$, we had to extrapolate the
continuum measured to the wavelength required using the fitted power law, but
given the fact that the continuum is well characterized in the $Z$-- and
$J$--band, this should not introduce large errors.  The black hole masses for
the sample at hand range from 0.3 to 5.2 $\times$ 10$^9$ M$_\odot$, including
the lowest black hole masses ever observed for $z \sim 6$ QSOs.  In general,
the black hole masses derived using the three different techniques agree
within a factor of a few, where the mass derived from the \ion{C}{4} line is
largest and the Eddington mass in between.  We note that, even in the absence
of any measurement errors, the individual BH mass estimates have an
uncertainty of a factor $\sim 3$ from the scatter in the correlations measured
by \citet{mcl04} and \citet{ves06}.

If the quasars were strongly lensed, the mass derived by the Eddington
luminosity would be larger by $\sqrt{q}$, where $q$ is the lensing factor, as
$M_{\rm Edd} \sim q$ and $M_{\rm MgII/CIV}$ roughly scales with $\sqrt{q}$).
Assuming that quasars emit isotropically \citep[which may not be the case,
see][]{hen07}, the good agreement between the different black hole mass
estimates implies that the quasars are likely not strongly lensed, consistent
with the conclusions derived from an analysis of HST images of these objects,
none of the QSOs showing evidence for lensed counterparts \citep{ric04,ric06a}.
For J1030+0524, \citet{hai02} predict a minimum BH mass of $\sim 10^9 M_\odot$
based on the flux distribution of the \lya\ emission alone and assuming it is
not gravitationally lensed.  The BH mass derived by us for this quasar is
in the range $1.4-3.2\,10^9 M_\odot$, consistent with the prediction by
\citet{hai02}.

\begin{deluxetable}{lrrrr}
\tabletypesize{\scriptsize}
\tablecaption{Estimated black hole masses\label{tbl:bh_masses}}
\tablewidth{0pt}
\tablehead{
\colhead{QSO} & \colhead{$z$\tablenotemark{a}} & 
\colhead{$M_{\rm BH}$(Mg\,{\scriptsize II})} &
\colhead{$M_{\rm BH}$(C\,{\scriptsize IV})} &
\colhead{$M_{\rm BH}$(Edd)} \\
&& \colhead{[10$^9$ M$_\odot$]} & \colhead{[10$^9$ M$_\odot$]} &
\colhead{[10$^9$ M$_\odot$]} 
}
\startdata
J0836+0054   & 5.82 & 2.7$\pm$0.6 &             & 4.0 \\
J0005$-$0006 & 5.85 & 0.3$\pm$0.1 & 0.7$\pm$0.1 & 0.7 \\
J1411+1217   & 5.95 & 1.1$\pm$0.1 & 1.2$\pm$0.2 & 1.2 \\
J1306+0356   & 5.99 & 2.4$\pm$0.4 & 5.2$\pm$0.7 & 1.9 \\
J1030+0524   & 6.28 & 1.4$\pm$0.2 & 3.2$\pm$0.6 & 1.8   
\enddata
\tablenotetext{a}{The redshift as published in the discovery paper.}  
\end{deluxetable}

Black hole masses derived from the observed--frame NIR emission lines for the
QSOs in our sample have not been reported in the literature, although for some
of the sources the \ion{C}{4} line has been observed (J0836+0054 by
\citealt{ste03}, J1306+0356 by \citealt{mai04a} and \citealt{pen02},
J1030+0524 by \citealt{mai04a}).  We can compare, however, with the BH mass
measured for the $z = 6.4$ QSO J1148+5251, which was measured by \citet{bar03}
and \citet{wil03}, employing relations between the \ion{Mg}{2} and \ion{C}{4}
line widths and black hole masses very similar to those employed by us.  These
authors find a BH mass of $(2-6)\times 10^9$ M$_\odot$ and $3\times 10^9$
M$_\odot$, respectively.  The BH masses measured by us are in general
smaller, showing that we sample the BH mass function to lower masses than
previous studies, probing not only the most massive and luminous objects but
also the presumably more common less massive quasars.

\subsection{Redshifts of QSOs in our sample}
\label{ssec:redshifts}

For several reasons, it is useful to know the systemic redshift of a QSO with
high precision, e.g., for follow--up observations of molecular lines with
narrow bandpasses.  The systemic redshift is difficult to derive from the
centroid of the observed \lya\ line, as it is very susceptible to absorption
by neutral hydrogen on its blue side, can be blended with the \ion{N}{5} line
or because it is completely absent.  Narrow emission lines due to star
formation are more suitable redshift indicators.  For example, the
[\ion{O}{3}]$\lambda$5007 line is presumed to be at nearly the systematic
redshift of the host galaxy \citep[see Sec.\ 3.0 of][for references]{vdb01}.
If narrow lines cannot be observed, broad QSO emission lines other than \lya\
can give important information about the systemic redshift of the galaxy
observed.  Using the redshift of [\ion{O}{3}]$\lambda$5007 as a zeropoint,
\citet{vdb01} find an average redshift offset of 143$\pm$91 and 161$\pm$10 km
s$^{-1}$ for \lya\ and \ion{Mg}{2} and a blueshift offset of 563$\pm$27 km
s$^{-1}$ for \ion{C}{4} (where the errors indicate uncertainties in the mean
value, the intrinsic scatter is much larger).

\citet{ric02b} study these offsets in more detail and find a redshift offset
of 97 km s$^{-1}$ for \ion{Mg}{2} and a blueshift offset of 727 km s$^{-1}$
for \ion{C}{4}, with dispersions of $\pm$269 and $\pm$511 km s$^{-1}$,
respectively.  Their analysis shows that the apparent shift of the \ion{C}{4}
line peak is not so much a shift as it is a lack of flux in the red wing.
They also confirm the observed anticorrelation between the shift of the
\ion{C}{4} line peak and the rest equivalent width of this line.
\citeauthor{ric02b}\ present this correlation as follows: they first sorted
their list of 794 quasars according to their \ion{C}{4}$-$\ion{Mg}{2} redshift
differences and then divided this list into four roughly equal bins with
$\sim$200 quasars each.  The four bins, having \ion{C}{4} EW$_0$s of 30.3,
25.8, 22.3, 18.0\,\AA, have average blue shifts of 193, 651, 1039, 1596 km
s$^{-1}$ with respect to the \ion{Mg}{2} line peak, respectively.  As all our
measured \ion{C}{4} EW$_0$s are larger than 30\,\AA, we assume the \ion{C}{4}
line is blue shifted by 96 km s$^{-1}$ with respect to the systemic velocity.
As the shifts of line peaks of emission lines in the spectrum of an individual
quasar are correlated, one can obtain a better redshift correction if the peak
wavelengths of several lines are measured \citep[see][]{she07}, but we do not
apply this method as we do not have enough suitable lines with a high enough
signal--to--noise ratio in our spectra.

\begin{deluxetable}{lrrrr}
\tabletypesize{\scriptsize}
\tablecaption{Redshift measurements\label{tbl:redshifts}}
\tablewidth{0pt}
\tablehead{
\colhead{QSO} & \colhead{$z$\tablenotemark{a}} & 
\colhead{$z_{\rm C\,IV}$} &
\colhead{$z_{\rm Mg\,II}$} &
\colhead{$z_{\rm sys}$\tablenotemark{b}}
}
\startdata
J0836+0054  & 5.82 &    --           & 5.810$\pm$0.003 & 5.808$\pm$0.007 \\
J0005$-$0006& 5.85 & 5.848$\pm$0.001 & 5.850$\pm$0.003 & 5.848$\pm$0.007 \\
J1411+1217  & 5.95 & 5.911$\pm$0.001 & 5.904$\pm$0.002 & 5.902$\pm$0.007 \\
J1306+0356  & 5.99 & 5.997$\pm$0.001 & 6.016$\pm$0.002 & 6.014$\pm$0.007 \\
J1030+0524  & 6.28 & 6.262$\pm$0.003 & 6.308$\pm$0.001 & 6.306$\pm$0.006 \\
\tableline
$\delta v$\tablenotemark{c} (km s$^{-1}$) && -193 & 97\\
$\delta z$\tablenotemark{d} && 0.005 & -0.002
\enddata
\tablenotetext{a}{The redshift as published in the discovery paper.}
\tablenotetext{b}{Systemic redshift estimate based on corrected Mg\,II
  redshift.}  
\tablenotetext{c}{Velocity difference between the listed line
  and the [O\,III]$\lambda$5007 centroid (from \citealt{vdb01} for \lya\ and
  N\,V, from \citealt{ric02b} for C\,IV and Mg\,II).}
\tablenotetext{d}{Correction in redshift for $z = 6.0$.}
\end{deluxetable}

Table \ref{tbl:redshifts} presents for each QSO, the observed \ion{C}{4} and
\ion{Mg}{2} redshift and the systemic redshift $z_{\rm sys}$, which is equal
to the \ion{Mg}{2} redshift corrected as described above (i.e., +97
km\,s$^{-1}$).  The error on the systemic redshift takes into account the
dispersion found by \citet{ric02b} for the offset of \ion{Mg}{2} (i.e.,
$\pm$269 km\,s$^{-1}$).  Relatively large inconsistencies remain between the
corrected redshifts of \ion{C}{4} and \ion{Mg}{2}.  One cause could be that
\ion{C}{4}, being a resonant line, can potentially suffer from absorption, as
is the case for J1411+1217 (see Fig.~\ref{fig:civ_plots}).

\vspace{1ex}
\centerline{\emph{Notes on individual objects}}
\vspace{1ex}

\noindent {\bf J0836+0054} This QSO was observed by \citet{ste03}. From the
\ion{C}{3}$\lambda$1909 line, \citet{ste03} derive a corrected\footnote{Using
  the 224 km s$^{-1}$ velocity difference with [O\,III]$\lambda$5007 found by
  \citet{vdb01}} redshift of $z = 5.774\pm0.003$.  This is quite different
from our systemic redshift of $z_{\rm sys} = 5.808\pm0.007$.  \citet{fre03}
measure an uncorrected \ion{C}{3}]$\lambda$1909 and \ion{Mg}{2} redshift of $z
= 5.82$ from their low resolution NICMOS spectrum.\smallskip

\noindent {\bf J1306+0356} Measurements to the same data presented here were
published by \citet{pen02}, who find an uncorrected \ion{C}{4} redshift of $z
= 6.00\pm0.01$, consistent with our result and with the value derived fom
\ion{N}{5} by \citet{bec01}.\smallskip

\noindent {\bf J1030+0524} \citet{fre03} measure an uncorrected
\ion{C}{3}]$\lambda$1909 and \ion{Mg}{2}$\lambda$2800 redshift of $z = 6.28$
from their low resolution NICMOS spectrum.  \citet{iwa04} find an uncorrected
\ion{Mg}{2} redshift of $z = 6.311$ (and do not report the uncertainty).
\citet{pen02} derive uncorrected redshifts of $z = 6.29\pm0.01$ and
$6.27\pm0.02$ from \ion{N}{5} and \ion{C}{4}, respectively, while we find a
consistent redshift of $z = 6.262\pm0.003$ (uncorrected) from \ion{C}{4}.

\subsection{The size of the ionized sphere around the QSO}

\begin{figure}
\includegraphics[angle=00,width=\columnwidth]{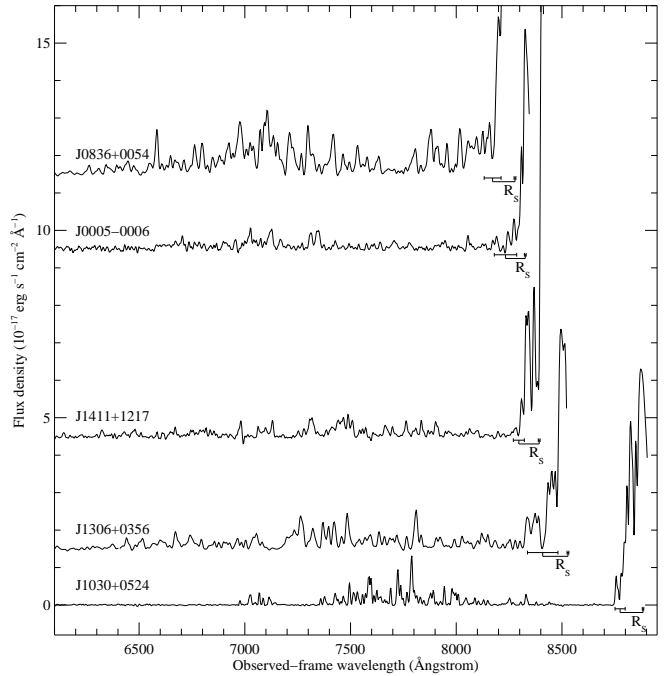}
\caption{Optical spectra of the quasars in our sample, offset in the vertical
  direction for clarity by 1.5, 4.5, 9.5, and 11.5 
  $\times 10^{-17}$\,erg\,s$^{-1}$\,cm$^{-2}$\,\AA$^{-1}$.  The size of the 
  Str\"omgren sphere ($R_{\rm S}$) is indicated by the line which 
  connects the measurements of the
  onset of the Gunn--Peterson (GP) trough ($z_{\rm GP}$) and the systemic
  redshift ($z_{\rm sys}$) derived
  from the \mgt\ line.  Error bars for both the GP trough onset and the
  systemic redshift have been indicated, although the uncertainties on the
  latter are too small for the error bars to be
  discernable for most quasars.
  \label{fig:GP_troughs}}
\end{figure}

At $z \sim 6$, most of a quasar's emission below the Ly$\alpha$ line is
absorbed by neutral hydrogen in the IGM.  However, the quasar is able to
ionize a (Str\"omgren) sphere around it, from which it transmits radiation at
wavelengths below that of the \lya\ line.  The red edge of the GP trough marks
the radius of this sphere.  Therefore, the radius of the Str\"omgren sphere
can be derived from the difference between the redshift of this red edge and
the systemic redshift.  However, if the GP effect is not complete, the precise
redshift $z_{\rm GP}$ of the red edge may be difficult to determine and may
depend on the spectral resolution of the observations.  We have tried to
determine $z_{\rm GP}$ conservatively by eye, determining where the \lya\
emission from the QSO may be fully absorbed, but including in its error the
range of redshifts where this plausibly could happen.  This results in
relatively large error bars for some QSOs in our sample, as shown in
Fig.~\ref{fig:GP_troughs}.  As the uncertainty in $z_{\rm GP}$ is much larger
than the uncertainty in the systemic redshift, we disregard the latter in the
following computation.

The Str\"omgren sphere radius ($R_{\rm s}$) is equal to the physical distance
between $z_{\rm sys}$ and $z_{\rm GP}$ \citep[see][]{hog99}:
\begin{equation}
R_{\rm S} = {D_{\rm H} \over (1+z)} \int^{z_{\rm sys}}_{z_{\rm GP}} 
(\Omega_{\rm m}(1+z)^3 + \Omega_\Lambda)^{-1/2} \,dz,
\end{equation}
where $D_{\rm H}$ is the Hubble distance and $\Omega_{\rm m}$ and
$\Omega_\Lambda$ the ratios of the mass and vacuum energy density to the
critical density, respectively.  For the cosmological parameters used here and
a redshift $z = 6.0$, we obtain $R_{\rm S} = 60 \, (z_{\rm sys} - z_{\rm GP})
~ {\rm Mpc}$.  The derived values of $R_{\rm s}$ for our sample are listed in
Table~\ref{tbl:stroemgren}.

\begin{deluxetable*}{lrrrrrrrr}
\tabletypesize{\scriptsize}
\tablecaption{Str\"omgren sphere radii and quasar activity time scales
\label{tbl:stroemgren}}
\tablewidth{0pt}
\tablehead{
\colhead{QSO} & \colhead{$z$\tablenotemark{a}} & 
\colhead{$M_{1450}$\tablenotemark{b}} &
\colhead{$z_{\rm GP}$} &
\colhead{$z_{\rm sys}$} &
\colhead{$R_{\rm S}$} &
\colhead{$\dot{N}_{\rm ph,Elv}$} &
\colhead{$t_{\rm Q,Elv}$} &
\colhead{$t_{\rm Q,Tel}$} \\
&&&&& \colhead{[Mpc]} & \colhead{[10$^{57}$ s$^{-1}$]} & 
\colhead{[10$^7$ yr]} & \colhead{[10$^7$ yr]}
}
\startdata
J0836+0054  &5.82&-27.83&5.721$\pm$0.033&5.808&5.7$\pm$2.2&4.70&0.6& 0.1 \\
J0005$-$0006&5.85&-26.42&5.771$\pm$0.044&5.848&4.9$\pm$2.8&0.47&3.9& 0.8 \\
J1411+1217  &5.95&-26.70&5.823$\pm$0.021&5.902&5.0$\pm$1.4&1.07&1.8& 0.4 \\
J1306+0356  &5.99&-27.14&5.916$\pm$0.059&6.014&5.9$\pm$3.6&1.40&2.4& 0.5 \\
J1030+0524  &6.28&-27.10&6.217$\pm$0.020&6.306&4.9$\pm$1.1&1.35&1.4& 0.3
\enddata
\tablenotetext{a}{The redshift as published in the discovery paper.}
\tablenotetext{b}{AB magnitude at rest--frame wavelength 1450\,\AA\ from the
  discovery papers, corrected for the cosmology used here.}
\tablenotetext{c}{Redshift from the centroids of the CO lines measured by
  \citealt{wal03} and \citealt{ber03a}.}
\end{deluxetable*}

\subsection{QSO activity time scales}

The size of \ion{H}{2} regions around a luminous quasar, prior to complete
reionization, depends on the ionizing photon emission rate ($\dot{N}_{\rm
  ph}$), the time span the ionization is sustained ($t_{\rm Q}$) and the
density of neutral hydrogen ($n_{\rm H} x_{\rm HI}$) in which the sphere expands
\begin{equation}
R_{\rm S} \approx \left( {3\, \dot{N}_{\rm ph}\, t_{\rm Q} \over 4 \pi\, n_{\rm H}\,
x_{\rm HI}} \right) ^{1/3}
\end{equation}
where $x_{\rm HI}$ is the neutral hydrogen fraction.  This formula results
from matching the total number of ionizing photons emitted to the number of
hydrogen atoms within a sphere, and it ignores both recombinations and the
Hubble expansion \citep[][see \citealt{mas07} for a discussion of the validity
of the assumptions made here]{whi03}.  If the quasar is located in a region
overdense by a factor $D$, the radius is reduced by $D^{-1/3}$ \citep{whi03}.
Furthermore, from the analysis of mock spectra along lines of sight through a
simulated QSO environment, \citet{mas07} find that the \ion{H}{2} region size
derived from quasar spectra is on average 30\% smaller than the physical one.
Note that issues of geometry, \ion{H}{1} clumping factor, and pre--QSO
reionization by, e.g., star forming galaxies in the QSO vicinity, further
complicate a conversion from observed Str\"omgren sphere radius to quasar
activity time.  As we are interested only in order of magnitude effects, we do
not take these issues into account.

Assuming that the quasar is located in a completely neutral IGM (i.e., $x_{\rm
  HI}\sim1$) which contains nearly all the baryons (i.e., $n_{\rm H}$ is equal
to the mean cosmic density at $z = 6$), \citet{cen00} and \citet{hai02} use
the the radius of the Str\"omgren sphere to constrain the minimum activity
time for the quasar
\begin{equation}\label{eq:ionization_time}
t_{\rm Q} = 
\left ({\dot{N}_{\rm ph}} \over {10^{57}\, {\rm s^{-1}}} \right )^{-1} 
\left ( {R_{\rm S} \over 7.4\, {\rm Mpc}} \right )^3 
\left ( {1 + z \over 7.0}                \right )^3 
\, 10^8 \, {\rm yr}.
\end{equation}

Note that this time is very short, and of the order of the light
crossing time of the Str\"omgren sphere.  \citet{whi03} explain that the speed
of light drops out of the equations describing the early evolution of a
Str\"omgren sphere, and that, for an observer, there is an initial period of
superluminal expansion.

The ionization rate depends on the SED below (blueward of) the Ly$\alpha$
line, which we estimated from the SED above 1216\,\AA.  The composite quasar
SED of \citet{elv94} with median bolometric luminosity 10$^{44}$ erg s$^{-1}$
has a corresponding median $\dot{N}_{\rm ph} = 5.5 \times 10^{53}$ photons
s$^{-1}$. We have used this relation to derive the ionization rates for our
sample QSOs ($\dot{N}_{\rm ph,Elv}$, see Tab.~\ref{tbl:stroemgren}).  From
these ionization rates and Eq.~\ref{eq:ionization_time} we derive minimum
activity times for the quasars in the range: $t_{\rm Q,Elv} \sim 0.6-9 \times
10^7$\,yr. However, as noted by \citet{whi03}, who have used the quasar
templates by \citet{tel02}, the ionization rates by \citeauthor{elv94}\ may be
too low by a factor of five.  Consequently, the quasar activity times derived
from the Str\"omgren spheres would go down by the same factor: $t_{\rm
  Q,Tel}\sim 0.1-2 \times 10^7$\,yr (see last column of
Tab.~\ref{tbl:stroemgren}).  For a partly \citep[$x_{\rm HI} = 0.1$,][]{wyi05}
or even mostly \citep[$x_{\rm HI} = 0.01$,][]{mas07} ionized IGM at $z \sim
6$, the derived activity times are smaller by a factor 10 or even 100.

We can now compare these ages to the e--folding time scale for the central
accreting black hole \citep[the \emph{Salpeter time scale},][]{sal64}
\begin{equation}
  t_{\rm acc} = 4\times10^7\, (\frac{\epsilon}{0.1}) \,\eta^{-1} \, {\rm yr}
\end{equation}
where $\epsilon$ is the radiative efficiency and $\eta$ is the ratio of quasar
to Eddington luminosity \citep[][see also the discussion in
\citealt{whi03}]{hai01}.
Assuming that the quasar is radiating at the Eddington limit ($\eta \sim 1$)
and an efficiency of $\epsilon \sim 0.1$, this simple calculation gives values
that are consistent, to first order, with our derived ages of the Str\"omgren
spheres in a neutral IGM.  This implies that the mass of the central black
hole in the quasars observed grows only by one e--folding or less during one
quasar activity cycle.

\subsection{Absorption features}

No Broad Absorption Lines (BALs) are found in the spectra, although the
\ion{C}{4} emission line of J1411+1217 shows associated \ion{C}{4} absorption.
The blue shifted BALs in QSOs are probably caused by an outflowing wind. At
low and intermediate redshift, BAL QSOs account for $\sim$15\% of the whole
quasar population \citep{rei03,hew03}, but \citet{mai04a} find four BAL QSOs
among the eight QSOs at $4.9 < z < 6.4$ that they observe, suggesting that the
most distant quasars are surrounded by a much larger amount of dense gas than
lower redshift quasars.  Adding the three QSOs without \ion{C}{4} BALs of our
sample (J1030+0524 was already found to be non--BAL by \citeauthor{mai04a}),
the fraction of BAL QSOs at $z > 4.9$ (5.7) goes down to 36\% (25\%).  A
larger sample of $z \sim 6$ QSOs with \ion{C}{4} spectroscopy is needed to
resolve this issue.

\begin{figure}
\includegraphics[angle=90,width=\columnwidth]{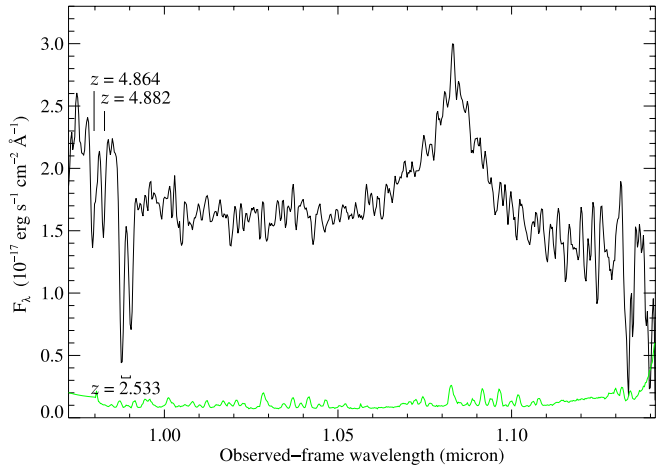}
\caption{$Z$--band spectrum of J1306+0356, smoothed over three pixels, with 
  Al\,{\scriptsize II} absorbtion from systems at $z = 4.864$ and 4.882 and
  \mgt\ absorbtion at $z = 2.533$ indicated.  The grey [green] line 
  shows the noise per pixel. [See the
  electronic edition of the Journal for a color version of this figure.]
  \label{fig:Z1306}}
\end{figure}

While not immediately the science focus of this paper, we have also searched
for narrow absorption lines of intervening systems.  The spectral resolution
of $\sim 20$\,\AA\ in our spectra does not allow to identify the \ion{C}{4}
doublet \citep[see][]{sim06,web06}, but it is good enough to resolve the
redshifted lines of the \ion{Mg}{2}$\lambda$2796,2803 system.  Within the
continuum emission of the QSO J1306+0356, we find two systems of double
absorption lines: at 9797.7, 9827.4\,\AA\ and 9880.7, 9906.3\,\AA\ with
observed EWs of 6.2, 4.0, and 11.6, 10.3\,\AA, respectively.  In addition, a
third system of double absorption may be present at the red edge of the
$J$--band spectrum: at 11340.6, 11401.2\,\AA\ with observed EWs of 13,
11\,\AA, respectively (see Fig.~\ref{fig:Z1306}).  These absorption features
would be consistent with \ion{Mg}{2} absorption at $z = 2.504$, 2.533 and
3.060 (although the latter only marginally), but a systematic search with
larger wavelength coverage by Jiang et al.\ (2007, AJ, in press) shows that
the correct identifications for the first double absorption feature is
\ion{Al}{2}\,1670 at $z = 4.864$ and 4.882, while the third double absorption
feature may be \ion{Fe}{3}\,1926 from the same intervening systems.  We note,
however, that the latter lines are at the edge of the observed spectrum where
the noise is higher and the wavelength calibration is not very accurate.  The
second double absorption feature is indeed confirmed to be \ion{Mg}{2} at $z =
2.533$ from other absorption lines identified.  The two absorbers at $z = 4.9$
are the highest redshift \ion{Mg}{2} absorbers known \citep[see,
e.g.,][]{kob05}.  See Jiang et al. for more details and a discussion on the
absorber host galaxies.

\subsection{Continuum slope}

The continuum slopes are measured over small wavelength intervals with little
pure continuum emission, especially for the $K$--band spectra.  Despite
resulting uncertainties, the measured slopes lie in the range $-2.06 < \alpha
< -1.83$, steeper than that measured by \citet[][-1.54]{vdb01} for the SDSS
composite QSO spectrum at a median $z = 1.3$ and than that measured from NIR
photometry of 45 QSOs at $3.6 < z < 5.3$ by \citet[][-1.43$\pm$0.33]{pen03}.
\citet{mai04a} find such steep continuum slopes for some of their QSOs at
$4.95 < z < 6.40$, and in particular for J1048+4637 at $z = 6.19$, where they
measure a slope of $\alpha = -2.1$ at rest frame wavelength above 1700\,\AA\
\citep{mai04b}.  Jiang et al. (2007, AJ, in press) observe a $z \sim 6$ QSO
sample with larger wavelength coverage and are therefore able to constrain the
continuum slopes better, obtaining results similar to ours.


\section{Summary and conclusions}\label{sec:summary}

We present sensitive near--infrared spectroscopic observations of a sample of
five $z \sim 6$ quasars, comprising all published quasars at $z > 5.8$
accessible from the VLT.  Our ISAAC spectra cover the \ion{C}{4}, \ion{Mg}{2}
and \ion{Fe}{2} lines, which are powerful probes of the chemical enrichment
and the black hole masses in these objects.  As the \ion{Mg}{2} and
\ion{Fe}{2} lines are blended (and are located on top of the Balmer
continuum), we have used Fe templates derived from an SDSS composite quasar
spectrum \citep{vdb01} and from the spectrum of I Zw 1 \citep{ves01} and
simple models representing the continuum and Balmer pseudo--continuum to
decompose the spectra. We derive an average \ion{Fe}{2}/\ion{Mg}{2} (which is
a proxy for the Fe/$\alpha$) ratio of 2.7$\pm$0.8 for our sample, which is
similar to the values measured for quasar broad line regions at lower
redshifts.  We note that the values measured by us are lower by a factor of a
few than some previous measurements for individual quasars at $z > 5.8$
reported in the literature \citep{mai03,fre03,iwa04} but are consistent with
others \citep{bar03,fre03,iwa04}.  We attribute these differences to different
choices in the Fe templates used, the corresponding spectral decomposition and
the different quality of the spectra.  The lack of evolution in the observed
\ion{Fe}{2}/\ion{Mg}{2} ratio and the mere presence of iron in the $z \sim 6$
systems demonstrate that the quasars in our sample must have undergone a major
episode of iron enrichment in the short time span between the Big Bang and the
emission of radiation observed by us, i.e., in less than one Gyr, consistent
with the conclusions derived by previous studies \citep[see
also][]{bar03,mai03,fre03}.

We have derived central black hole masses using three different
methods: A) using the Eddington luminosity, B) using the empirical
equation relating the linewidth of \ion{Mg}{2} (and the nearby
continuum emission) to the mass of the black hole \citep{mcl04} and C)
using the corresponding method for the CIV line \citep{ves06}.  The
derived masses using the \ion{Mg}{2} and \ion{C}{4} lines and the
Eddington luminosity agree well, and we derive central black hole
masses of $0.3 - 5.2\times10^9$\,M$_\odot$.  Our derived black hole
mass range includes the lowest black hole mass ever measured at $z
\sim 6$ and indicates that we have started to probe the more typical
quasar population (with lower masses/luminosities) at this extreme
redshift.  Clearly, future, more sensitive, observations of an
extended sample towards lower luminosities are needed to constrain the
properties of the fainter quasar population further.  Assuming that
the quasars indeed radiate at their Eddington luminosity, the
agreement in derived black hole masses using the different methods
implies that the quasars are likely not strongly lensed and emit
roughly isotropically.

We use the difference between the redshift derived from the \ion{Mg}{2} line,
which serves as the best available proxy for the systemic redshift of the
quasar, and the redshift for the onset of the GP trough in the quasar spectra
to derive the extent of the ionized Str\"omgren spheres around our target
quasars.  The derived physical radii are relatively well constrained to about
five Mpc. Using simple ionization models, the central quasar activity would
need of order $10^6-10^8$\,yr to create these cavities, allowing the quasar
black hole to grow by one e--folding or less during an activity cycle.\\




\acknowledgments
JK is supported by the DFG, Sonderforschungsbereich (SFB) 439.  DR
acknowledges support from DFG Priority Programme 1177.  MAS acknowledges
support from NSF grant AST--0307409.  We are grateful to Marianne Vestergaard
for providing the electronic version of her \ion{Fe}{2} template
\citep{ves01}.  Based on observations carried out at the European Southern
Observatory, Paranal, Chile under program Nos.\ 267.A--5689, 069.B--0289,
071.B--0525, 074.A--0477 and 076.A--0304.



{\it Facilities:} \facility{VLT (ISAAC)}.




\bibliographystyle{apj}
\bibliography{ms} 

\end{document}